\documentclass[aps,prd,reprint,nofootinbib,preprintnumbers,floatfix,superscriptaddress]{revtex4-1}

\pdfoutput=1

\usepackage[bookmarks=false,hyperfootnotes=false]{hyperref}
\usepackage{graphicx}
\usepackage{amsmath}
\usepackage{subfig}
\usepackage{xspace}
\usepackage{slashed}
\usepackage{amssymb}

\usepackage{bbm}

\newcommand{\ecf}[2]{e_{#1}^{(#2)}} 
\newcommand{\ecfnobeta}[1]{e_{#1}}

\DeclareRobustCommand{\Sec}[1]{Sec.~\ref{#1}}
\DeclareRobustCommand{\Secs}[2]{Secs.~\ref{#1} and \ref{#2}}

\DeclareRobustCommand{\Tab}[1]{Table~\ref{#1}}

\DeclareRobustCommand{\Fig}[1]{Fig.~\ref{#1}}
\DeclareRobustCommand{\Figs}[2]{Figs.~\ref{#1} and \ref{#2}}
\DeclareRobustCommand{\Eq}[1]{Eq.~(\ref{#1})}
\DeclareRobustCommand{\Eqs}[2]{Eqs.~(\ref{#1}) and (\ref{#2})}
\DeclareRobustCommand{\Ref}[1]{Ref.~\cite{#1}}
\DeclareRobustCommand{\Refs}[1]{Refs.~\cite{#1}}

\newcommand{\Nsub}[2]{\tau_{#1}^{(#2)}}
\newcommand{\Nsubnobeta}[1]{\tau_{#1}}

\newcommand{\fastjet}[1]{\textsc{FastJet\xspace #1}}

\newcommand{\vincia}[1]{\textsc{Vincia\xspace #1}}
\newcommand{\nlojet}[1]{\textsc{NLOJet++\xspace #1}}

\begin{document}

\preprint{MIT--CTP 4725}

\title{
The Singular Behavior of Jet Substructure Observables
}

\author{Andrew J. Larkoski}
\email{larkoski@physics.harvard.edu}
\affiliation{Center for Fundamental Laws of Nature, Harvard University, Cambridge, MA 02138, USA}
\author{Ian Moult}
\email{ianmoult@mit.edu}
\affiliation{Center for Theoretical Physics, Massachusetts Institute of Technology, Cambridge, MA 02139, USA}

\begin{abstract}

Jet substructure observables play a central role at the Large Hadron Collider for identifying the boosted hadronic decay products of electroweak scale resonances. 
The complete description of these observables 
 requires understanding both the limit in which hard 
 substructure is resolved, as well as the limit of a jet with a single hard core. In this paper we study in detail the perturbative structure of two prominent jet substructure observables, $N$-subjettiness and the energy correlation functions, as measured on background QCD jets.
In particular, we 
focus on the distinction between the limits in which two-prong structure is resolved or unresolved.
Depending on the choice of subjet axes, we demonstrate that at fixed order, $N$-subjettiness can manifest myriad behaviors in the unresolved region: smooth tails, end-point singularities, or singularities in the physical region.
The energy correlation functions, by contrast, only have non-singular perturbative tails extending to the end point.
We discuss the effect of hadronization on the various observables with Monte Carlo simulation and demonstrate that the modeling of these effects with non-perturbative shape functions is highly dependent on the $N$-subjettiness axes definitions.
Our study illustrates those regions of phase space that must be controlled for high-precision jet substructure calculations, and emphasizes how such calculations can be facilitated by designing substructure observables with simple singular structures.

\end{abstract}

\maketitle

\section{Introduction}\label{sec:intro}

The identification of hadronically decaying boosted electroweak scale resonances for searches within and beyond the Standard Model is playing an increasingly important role as the Large Hadron Collider (LHC) resumes its operations at $13$ TeV. Significant theoretical \cite{Abdesselam:2010pt,Altheimer:2012mn,Altheimer:2013yza,Adams:2015hiv} and experimental \cite{Aad:2012meb,ATLAS:2012am,Aad:2013gja,Aad:2013fba,TheATLAScollaboration:2013tia,TheATLAScollaboration:2013sia,TheATLAScollaboration:2013ria,TheATLAScollaboration:2013pia,CMS:2013kfa,CMS:2013wea,CMS-PAS-JME-10-013,CMS-PAS-QCD-10-041,Aad:2014gea,CMS:2014joa,Aad:2014haa,atlas_recent:2015,Aad:2015rpa} effort has therefore been devoted to understanding jet observables capable of distinguishing the decay products of such resonances from the background of QCD jets. While such observables have been primarily studied with parton shower Monte Carlo generators, recent years have seen a significant advance in analytic calculations of jet substructure observables, and an understanding of their behavior to all orders in perturbation theory (see e.g. \Refs{Feige:2012vc,Field:2012rw,Dasgupta:2013ihk,Dasgupta:2013via,Larkoski:2014pca,Dasgupta:2015yua,Krohn:2012fg,Waalewijn:2012sv,Larkoski:2015kga}).

Of particular phenomenological importance are observables which are sensitive to hard two-prong substructure within a jet, relevant for tagging hadronically decaying boosted $W,Z,$ and Higgs bosons. Among the most widely applied observables are the $N$-subjettiness ratio observable $\Nsubnobeta{2,1}$ \cite{Thaler:2010tr,Thaler:2011gf} and ratio observables formed from the energy correlation functions \cite{Larkoski:2013eya}, namely $C_2$ \cite{Larkoski:2013eya} and $D_2$ \cite{Larkoski:2014gra}. Due to the important role these observables are playing at the LHC, it is essential that they be brought under theoretical control. As a first step in this direction, the $\Nsubnobeta{2,1}$ observable was calculated for boosted $Z$ jets in $e^+e^-$ collisions \cite{Feige:2012vc}. More recently an analytic calculation was performed for the $D_2$ observable for both QCD and $Z$ jets, also in $e^+e^-$ collisions \cite{Larkoski:2015kga}.

%%%%
\begin{figure*}[t]
\centering
\subfloat[]{
\includegraphics[width=13cm]{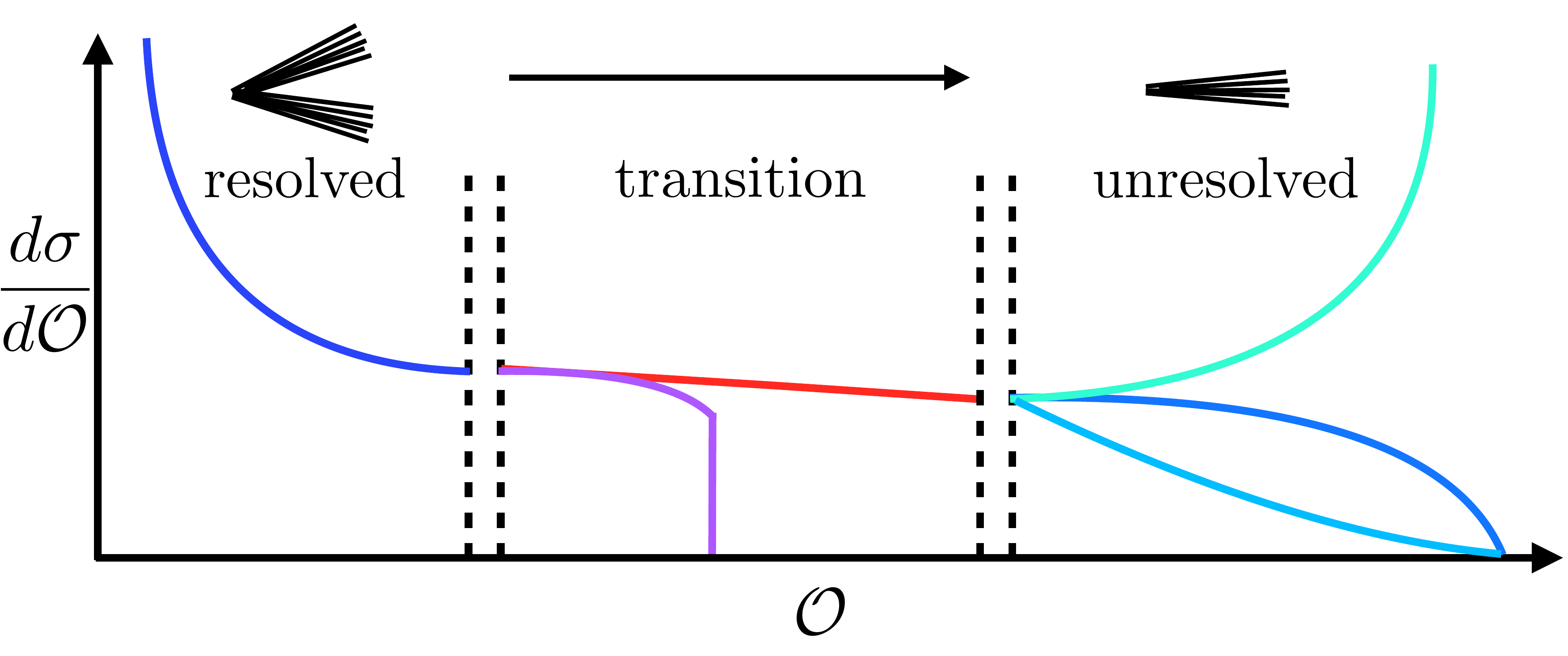}    
} 
\caption{The leading fixed order singular structure of a two-prong substructure observable, $\mathcal{O}$, showing different behavior in the resolved, transition, and unresolved regions. A prediction for $\frac{d\sigma}{d\mathcal{O}}$ requires a simultaneous understanding of each of the three regions.
}
\label{fig:sing_structure}
\end{figure*}
%%%%

Jet shape observables generically exhibit soft and collinear singularities, necessitating a resummation of singular contributions to all orders to achieve reliable predictions \cite{Kodaira:1981nh,Collins:1984kg,Sterman:1986aj,Catani:1989ne,Catani:1990rr,Catani:1991bd,Catani:1991kz,Catani:1992jc,Catani:1992ua,Catani:1996yz,Catani:1998sf}. Furthermore, they  receive large non-perturbative corrections associated with singular regions of phase space. A resummation of the singular contributions as well as a first principles treatment of non-perturbative effects can be achieved by proving an all-orders  factorization theorem describing a particular singular region of phase space. The resummed prediction can then be matched to a fixed order perturbative result valid away from the singular region. In the general case that a jet shape distribution exhibits multiple singular regions, distinct factorization theorems are required for each singular region.

While the resummation program is well understood for simple observables, for example $e^+e^-$ event shapes where it has been pushed to high orders \cite{deFlorian:2004mp,Becher:2008cf,Abbate:2010xh,Chien:2010kc,Becher:2012qc,Hoang:2014wka,Banfi:2014sua}, it has proven more complicated for jet substructure observables. The difficulty in obtaining a theoretical description of two-prong substructure observables is that they must be understood both in the limit of two resolved subjets, as well as in the limit that a subjet structure is not resolved. For example, when used for boosted boson discrimination, one is interested not only in the behavior when measured on a boosted $Z$ or $W$ jet, but also in the behavior when measured on QCD jets.

To understand how this complicates their theoretical description, consider a two-prong substructure observable, $\mathcal{O}$. As is typical of such observables, we will assume it is formed from a combination of two observables, one of which is first non-zero with a single emission off of the initiating parton and one of which is first non-zero with two emissions. Furthermore, we assume that $\mathcal{O}$ is chosen so that it identifies a two-prong structure in the limit $\mathcal{O}\to 0$, and satisfies $\mathcal{O}\gg 0$ when there is no resolved structure. A concrete example of such an observable is the $N$-subjettiness ratio observable $\Nsub{2,1}{\beta}$, which is constructed as the ratio of the $2$-subjettiness and $1$-subjettiness jet shape observables
%%%%
 \begin{align}\label{eq:obs_form}
 \Nsub{2,1}{\beta}=\frac{\Nsub{2}{\beta}}{\Nsub{1}{\beta}}\,,
 \end{align}
%%%%
each of which will be defined in \Sec{sec:obs}.

In fixed order perturbation theory the two-prong substructure observable, $\mathcal{O}$, which is sensitive to both single and double emissions,  can have singularities arising from distinct physical configurations. This is shown schematically in \Fig{fig:sing_structure}. In the limit $\mathcal{O}\to 0$, which we refer to as the resolved limit, the jet has a two prong substructure. In this region of phase space, $\mathcal{O}$ is set by a single soft or collinear emission off of the dipole structure of the hard splitting defining the substructure, and will generically exhibit a soft and collinear singularity. In the unresolved limit, where the jet does not exhibit a resolved substructure, a variety of different behaviors including smooth tails, kinematic endpoints, or endpoint singularities, are possible. Finally, there is a transition region, where discontinuities, or shoulders, can appear at fixed order in perturbation theory. This leads to large (possibly singular) corrections at higher orders, due a miscancellation of real and virtual corrections at the shoulder \cite{Catani:1997xc}.

A complete calculation of a two-prong substructure observable requires a description of the physics in each of the three regions indicated in \Fig{fig:sing_structure}, including a resummation of singular contributions and treatment of non-perturbative corrections in each region. Furthermore, the descriptions in the different regions must be matched. A first step towards this goal is a detailed understanding of the singular structure of the observable, which is the subject of this paper. 

In this paper we study the fixed order perturbative structure of two common substructure discriminants, namely the $N$-subjettiness observable, $\Nsubnobeta{2,1}$, and the ratio of energy correlation functions $D_2$. We show a number of interesting features regarding the singular structure of the $N$-subjettiness observable, which have not previously been discussed in the literature. For axes defined with an exclusive generalized $k_T$ algorithm \cite{Catani:1993hr,Cacciari:2008gp,Cacciari:2011ma}, we classify the singular behavior at the $\Nsubnobeta{2,1}\sim1$ endpoint, showing that the presence of an endpoint singularity depends on the choice of clustering metric. For axes defined through minimization, we show the presence of discontinuities in the physical region,  which necessitate the use of high order matrix elements to accurately describe the unresolved region. We contrast this behavior with the singular structure of the $D_2$ observable, which exhibits perturbative stability.

The outline of this paper is as follows. In \Sec{sec:obs} we define the $N$-subjettiness and energy correlation function observables, with a particular attention to the definition of the $N$-subjettiness axes.  In \Sec{sec:resolved} we study the behavior of the observables in the limit of two resolved subjets.  In \Secs{sec:endpoint}{sec:physical_region} we discuss the behavior of the $N$-subjettiness observable in the unresolved limit with generalized $k_T$ axes, and with axes defined through minimization. The behavior of the $D_2$ observable in the unresolved limit is studied in \Sec{sec:unresolved_D2}.  In \Sec{sec:non_pert} we discuss the impact of the singular structure of the observable on non-perturbative corrections due to hadronization. We conclude in \Sec{sec:conc}.

%%%%
\section{Observables}\label{sec:obs}
%%%%

A powerful class of observables for two-prong discrimination are those formed from ratios of either the $N$-subjettiness observables, or the energy correlation functions. In this section, we define these observables, as well as the  ratios commonly used in the study of jet substructure. 
For simplicity, throughout this paper we work with jets produced in $e^+e^-$ collisions, and we therefore give the definitions of the observables relevant for this case.

%%%%
\subsection{$N$-subjettiness}\label{sec:Nsub_def}
%%%%

We define the $N$-subjettiness observable, $\Nsub{N}{\beta}$ \cite{Stewart:2010tn,Thaler:2010tr,Thaler:2011gf} for the case of $e^+e^-$ colliders, as\footnote{$N$-jettiness has been defined and studied as an event shape in $e^+e^-$ in \Refs{Stewart:2010tn,Mateu:2012nk}. Their definition differs slightly from ours in the choice of normalization, as appropriate for an event shape as compared with a jet shape.} 
\begin{equation}\label{eq:Nsub_def}
\Nsub{N}{\beta} = \frac{1}{E_J}\sum_{i\in J} E_i \min\left\{
\frac{2 p_i\cdot n_1}{E_i},\dotsc,\frac{2 p_i\cdot  n_N}{E_i}
\right\}^{\beta/2}\,.
\end{equation}
Here $J$ denotes the jet, $E_i$ and $p_i$ are the energy and four momentum of particle $i$ in the jet,  $N>0$ is an integer defining the number of axes, $n_j$ are the lightlike vectors defining the directions of the axes, and $\beta$ is an angular exponent required to be greater than zero for infrared and collinear (IRC) safety. For notational simplicity, we will often drop the explicit angular exponent, denoting the observable simply as $\Nsubnobeta{N}$.

One subtlety in the definition of the $N$-subjettiness observable in \Eq{eq:Nsub_def} is the definition of the axes $n_i$. While their placement is clear in the limit of a resolved substructure,\footnote{The placement of axes is also clear when used as an event shape, namely $N$-jettiness \cite{Stewart:2010tn}, or the XCone algorithm \cite{Stewart:2015waa} with well separated jets. In these cases, factorization theorems in SCET have been proven.} an algorithmic definition is required to determine their behavior in the unresolved limit.  Two main approaches have been used for defining the axes. The first approach is to define the $N$-subjettiness axes as the axes found using an exclusive jet clustering algorithm. We will consider recursive clustering algorithms with the generalized $k_T$ metric \cite{Catani:1993hr,Cacciari:2008gp,Cacciari:2011ma}
%%%%
\begin{equation}\label{eq:gen_kt_metric}
\hspace{-0.025cm} d_{ij} = \min[E_i^{2p},E_j^{2p}]\frac{1-\cos\theta_{ij}}{1-\cos R}\sim  \min[E_i^{2p},E_j^{2p}]\frac{\theta_{ij}^2}{ R^2}\,.
\end{equation}
%%%%
Here $E_i$ denotes the energy of particle $i$, $\theta_{ij}$ denotes the angle between particles $i$ and $j$, and $p$ and $R$ are parameters which define the choice of metric. For numerical studies, we will focus on the exclusive Cambridge-Aachen (C/A, $p=0$) \cite{Dokshitzer:1997in,Wobisch:1998wt,Wobisch:2000dk}, exclusive $p=1/2$, and exclusive $k_T$ ($p=1$) \cite{Catani:1991hj} clustering algorithms. We will also discuss how the behavior is modified using winner-take-all (WTA) recombination  \cite{Bertolini:2013iqa,Larkoski:2014uqa,Larkoski:2014bia}, in contrast to traditional $E$-scheme recombination.

%%%%
\begin{figure*}[t]
\centering
\subfloat[]{\label{fig:ninja_pic}
\includegraphics[width=5cm]{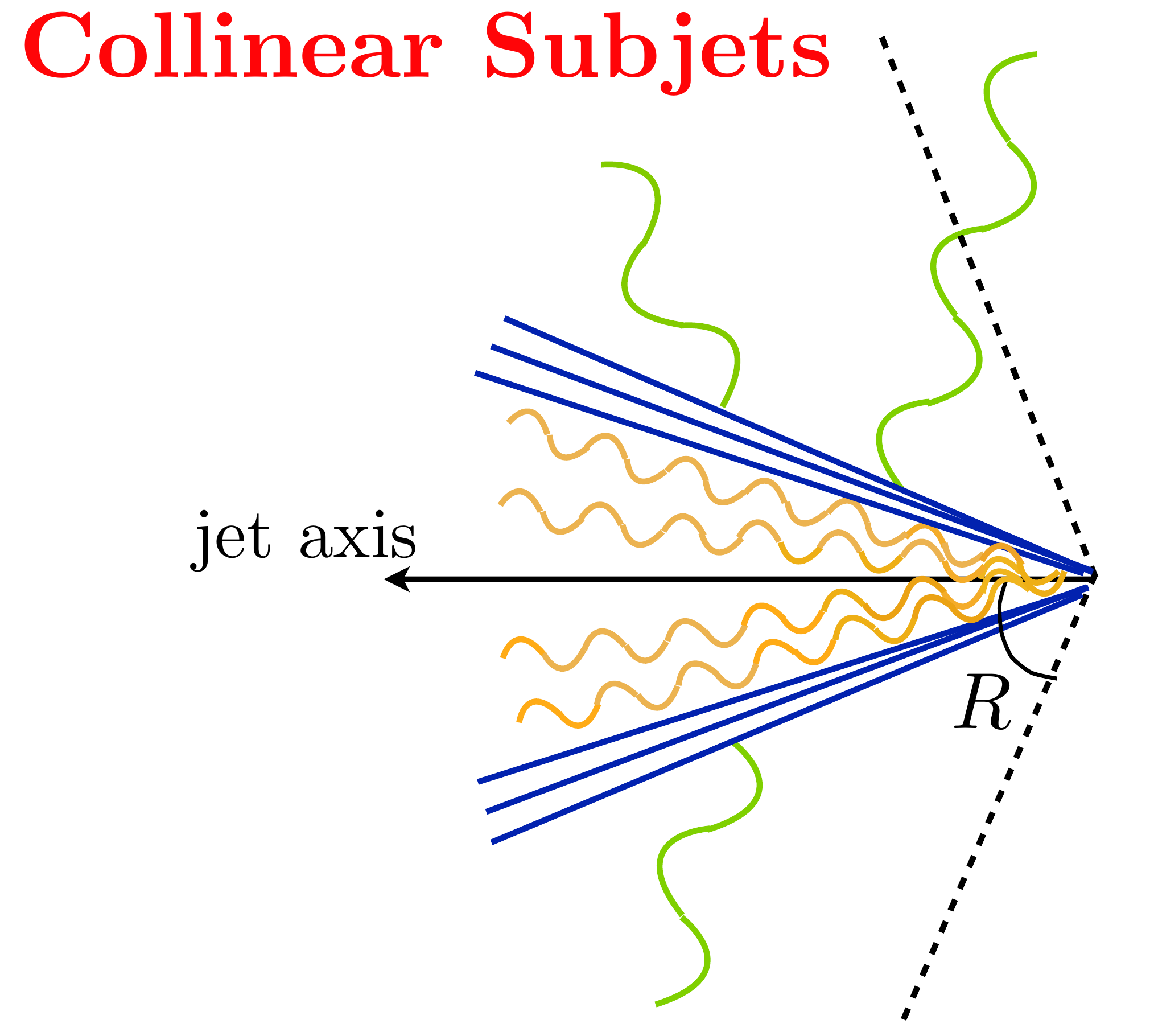}    
} \ \ \hspace{2.5cm}
\subfloat[]{\label{fig:ssj_pic}
\includegraphics[width=4.0cm]{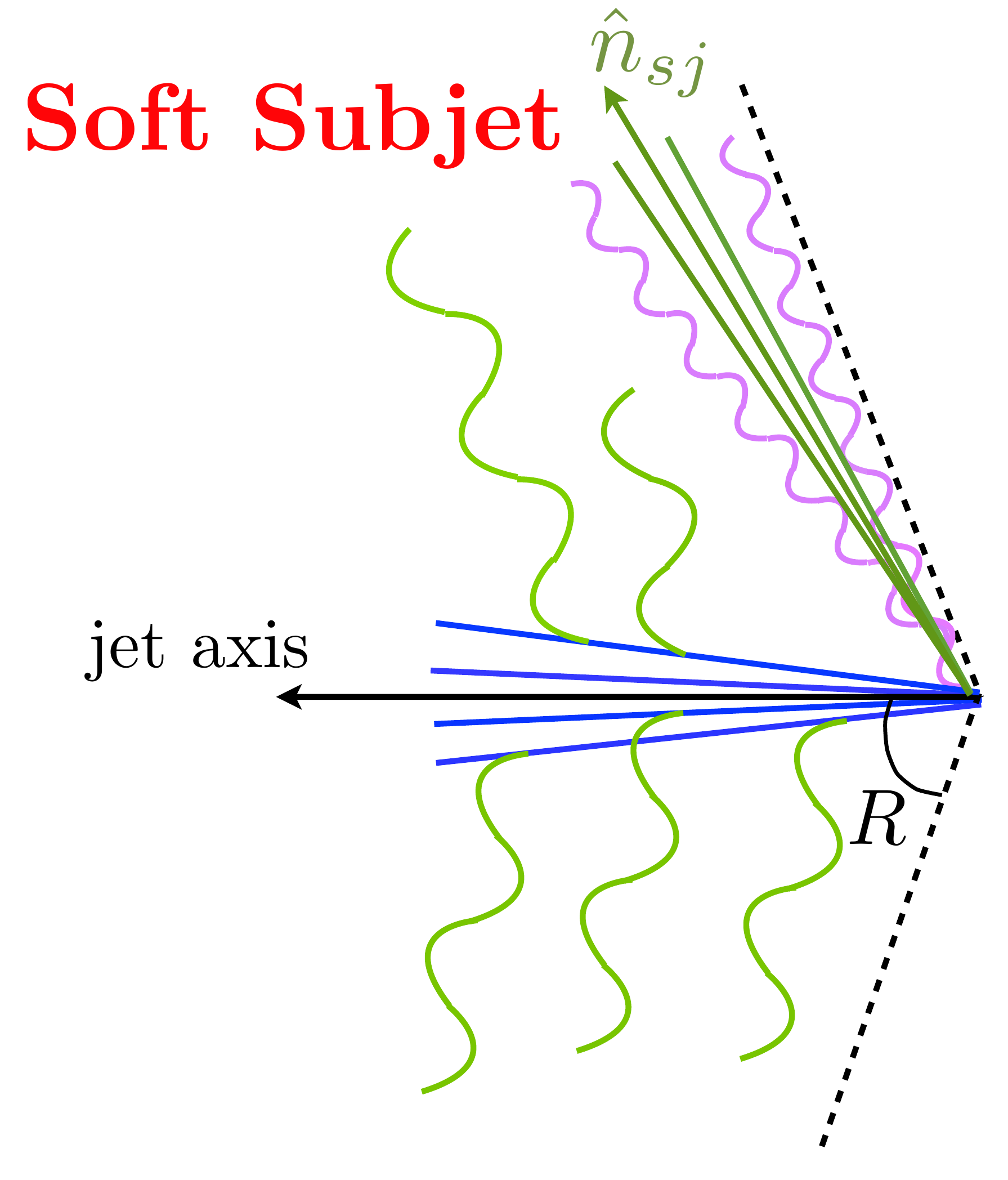} 
}
\caption{ A schematic depiction of the jet configurations contributing in the resolved limit. The collinear subjets configuration  is shown in (a) and the soft subjet configuration is shown in (b).  
}
\label{fig:jet_configs}
\end{figure*}
%%%%

The second approach to defining the $N$-subjettiness axes is to minimize the sum in \Eq{eq:Nsub_def} over possible light-like axes $n_i$. In practice, this minimization is performed starting from seed axes, typically defined using an exclusive jet clustering algorithm \cite{Thaler:2011gf,Stewart:2015waa}. In this case, it is essential that the seed axes are chosen optimally.  We use the approach of the XCone algorithm \cite{Stewart:2015waa,Thaler:2015xaa} which matches the parameters of the jet clustering metric and recombination scheme with the minimization metric of $N$-subjettiness. We find that this is essential to obtain sensible results, particularly in fixed order perturbation theory.\footnote{Multi-pass minimization can also be used, and is implemented in \fastjet{}. This procedure is non-deterministic, and not IRC safe.  Furthermore, a large number of passes are required to obtain reasonable results.} The particular clustering metric used for different values of the $N$-subjettiness exponent $\beta$ will be given in \Sec{sec:physical_region} when minimization is discussed.

$N$-subjettiness is primarily used in the form of a ratio observable. For tagging two prong substructure the appropriate ratio observable is \cite{Thaler:2010tr}
 \begin{align}
 \Nsub{2,1}{\beta}=\frac{\Nsub{2}{\beta}}{\Nsub{1}{\beta}}\,.
 \end{align}
This variable is Sudakov safe \cite{Larkoski:2013paa,Larkoski:2015lea} without a jet mass cut and is IRC safe with a jet mass cut. A mass cut is not equivalent to a cut on the $\Nsubnobeta{1}$ observable, which could also be used to make the observable IRC safe.  However, motivated by the application of boosted boson tagging where a narrow mass cut is applied, we choose to use a mass cut. 

The $ \Nsubnobeta{2,1}$ observable is of the form considered in \Sec{sec:intro} for the general observable $\mathcal{O}$, in particular, from \Eq{eq:Nsub_def} we see that $\Nsubnobeta{1}$ is set by a single emission from the parton initiating the jet, while $\Nsubnobeta{2}$ requires two emissions to be non-zero. From the definition of the $N$-subjettiness observable in \Eq{eq:Nsub_def}, we see that there exists a physical endpoint at $ \Nsubnobeta{2,1}=1$, which defines the unresolved limit. On the other hand, the resolved limit is defined by the relation  $\Nsubnobeta{2,1}\to 0$, so that the physical region for the observable is $ 0\leq \Nsubnobeta{2,1} \leq 1$.

%%%%
\subsection{Energy Correlation Functions}
%%%%

The (dimensionless) two and three point energy correlation functions are defined for the case of $e^+e^-$ colliders as \cite{Larkoski:2013eya}
\begin{align}\label{eq:ecf_def}
\ecf{2}{\beta}&= \frac{1}{E_J^2} \sum_{i<j\in J} E_i E_j \left(
\frac{2p_i \cdot p_j}{E_i E_j}
\right)^{\beta/2} \,, \\
\ecf{3}{\beta}&= \frac{1}{E_J^3} \sum_{i<j<k\in J} E_i E_j E_k \left(
\frac{2p_i \cdot p_j}{E_i E_j}
\frac{2p_i \cdot p_k}{E_i E_k}
\frac{2p_j \cdot p_k}{E_j E_k}
\right)^{\beta/2} \,. \nonumber
\end{align}
Here $J$ denotes the jet, $E_i$ and $p_i$ are the energy and four momentum of particle $i$ in the jet and $\beta$ is an angular exponent that is required to be greater than 0 for IRC safety. For notational simplicity, we will often drop the angular exponent $\beta$.

For tagging two-prong substructure, it has been shown that the appropriate variable is \cite{Larkoski:2014gra}\footnote{Another variable $C_2=\ecfnobeta{3}/\left(\ecfnobeta{2}\right)^2$ has also been proposed in \Ref{Larkoski:2013eya}. In the limit of a $\delta$-function mass cut, it is equivalent to $D_2$. The singular structure of the two observables is therefore closely related, so we restrict ourselves to considering $D_2$.}
%%%
\begin{align}\label{eq:def_D2}
D_2^{(\beta)}=\frac{\ecf{3}{\beta}}{\left(\ecf{2}{\beta}\right)^3}\,.
\end{align}
As with $\Nsubnobeta{2,1}$, $D_2$ is Sudakov safe, but is rendered IRC safe with a mass cut.
The $D_2$ observable is of the general form considered in \Sec{sec:intro} for the observable $\mathcal{O}$, namely from \Eq{eq:ecf_def} we see that the two point energy correlation functions are set by a single emission, while the three point correlation functions are first non-zero at two emissions.
This observable was factorized and resummed for both QCD jets and hadronically-decaying color singlets in \Ref{Larkoski:2015kga}.

The phase space for the $D_2$ observable has been studied in detail in \Refs{Larkoski:2014gra,Larkoski:2015kga}. The upper boundary of the phase space is found to scale as $\ecfnobeta{3} \sim ( \ecfnobeta{2}  )^2$, which defines the unresolved region. The resolved region of phase space is defined by $\ecfnobeta{3} \ll ( \ecfnobeta{2} )^3$, or $D_2 \ll 1$. A particular consequence of this phase space structure is a distinction between the behavior in the unresolved limit for $D_2$, as compared with $\Nsubnobeta{2,1}$. Without a mass cut, $D_2$ has no upper endpoint. In the presence of a mass cut, it can be shown that in the case of $\beta=2$ the physical endpoint can be written in terms of the jet energy, and jet mass, and is given by ${D_2^{(2)\,\text{max}}}=\frac{E_J^2}{m_J^2}\gg1$.

%%%%
\section{The Resolved Limit}\label{sec:resolved}
%%%%

%%%%
\begin{figure*}[t]
\centering
\subfloat[]{\label{fig:nsub_sb3}
\includegraphics[width=7.5cm]{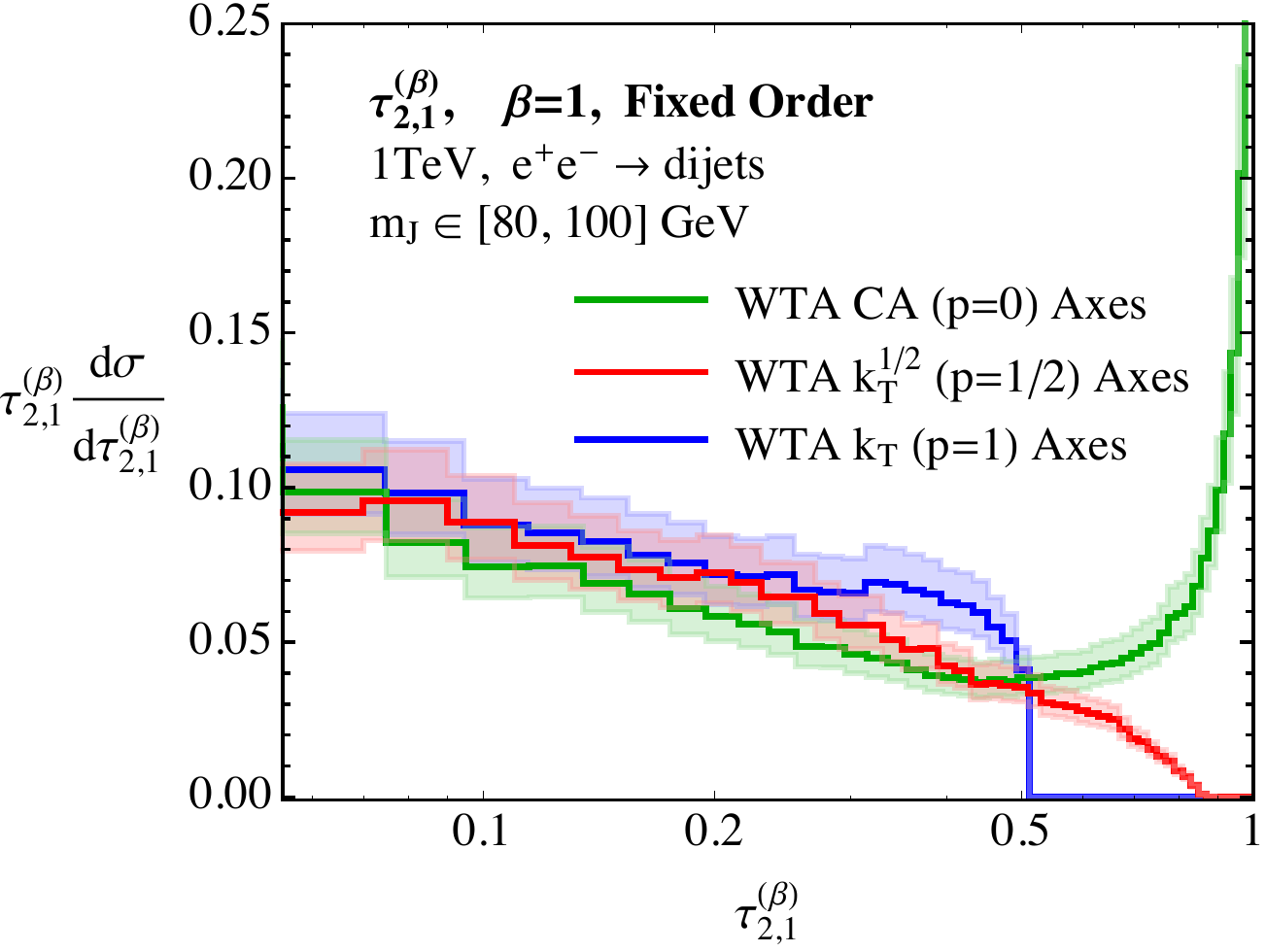}    
} \ \ \hspace{1cm}
\subfloat[]{\label{fig:nsub_sb4}
\includegraphics[width=7.5cm]{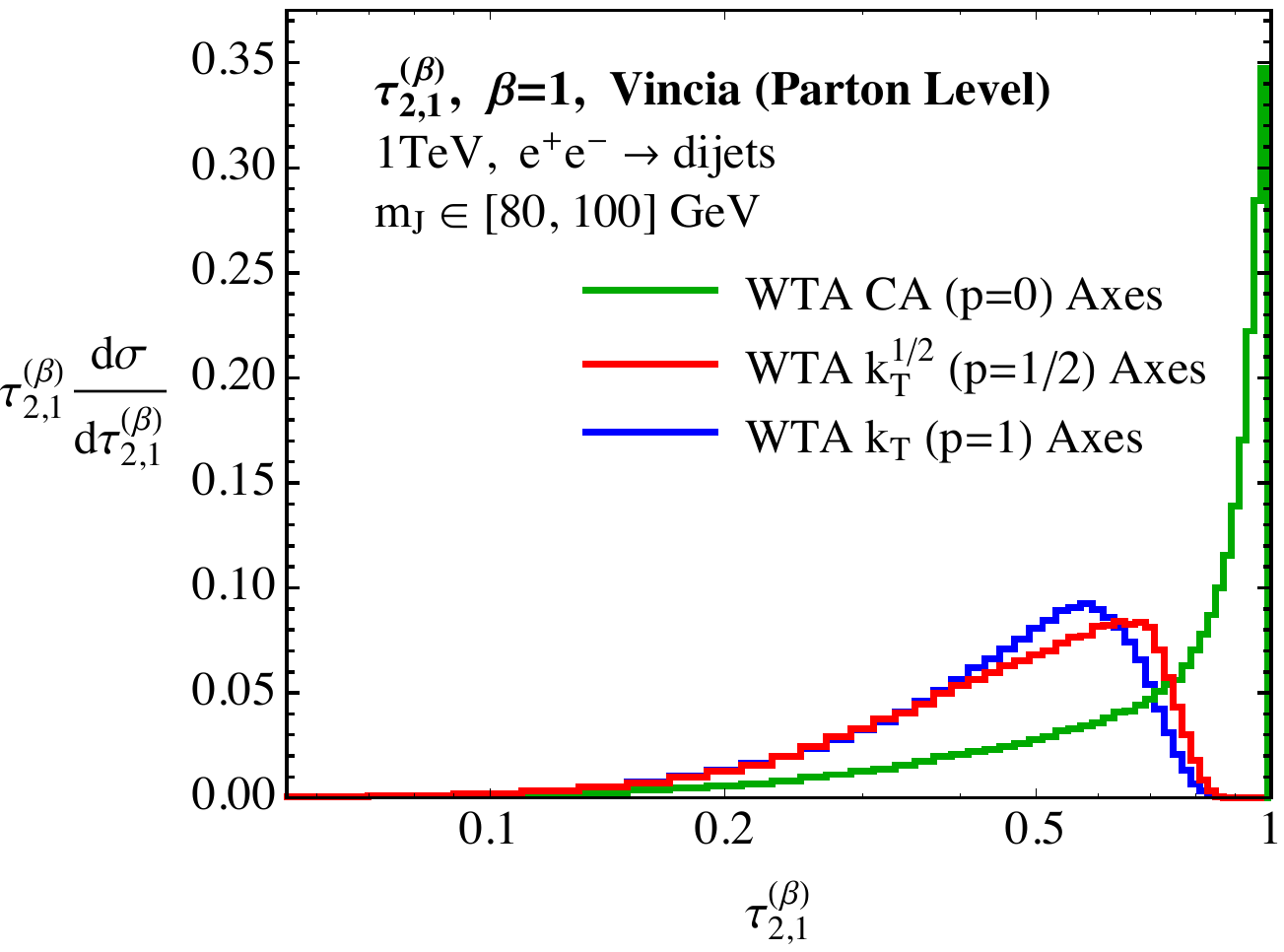} 
}\\
\subfloat[]{\label{fig:nsub_sb1}
\includegraphics[width=7.5cm]{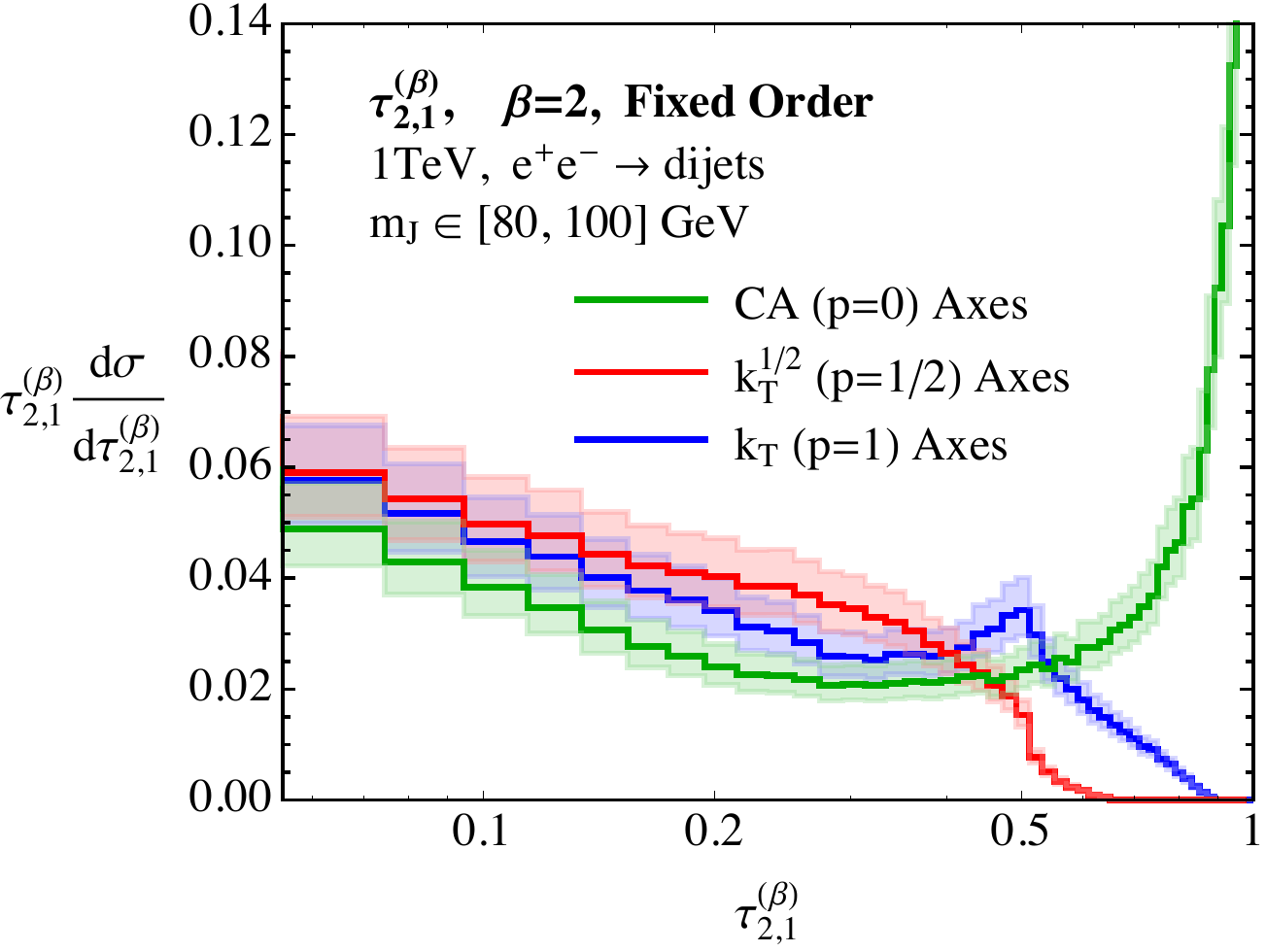}    
} \ \ \hspace{1cm}
\subfloat[]{\label{fig:nsub_sb2}
\includegraphics[width=7.4cm]{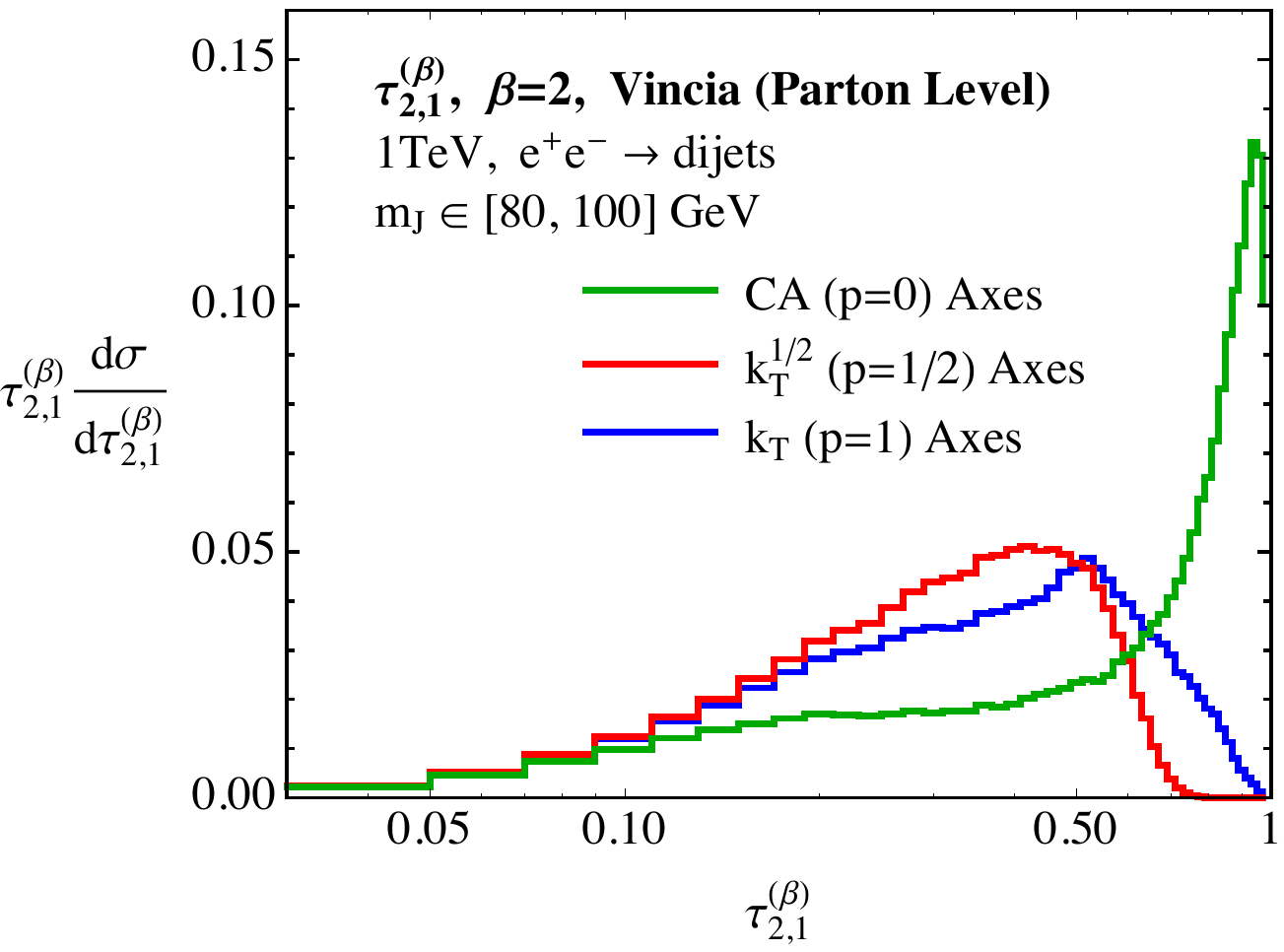} 
}\\
\caption{ The behavior of the $\Nsubnobeta{2,1}$ observable with $\beta=1$ and $\beta=2$ for different axes definitions in LO perturbation theory on the left, and in parton level \vincia{} Monte Carlo on the right. The universality observed in the $\Nsubnobeta{2,1}\to0$ limit should be contrasted with the different possible behaviors as $\Nsubnobeta{2,1}\to1$.
}
\label{fig:nsub_FO_2}
\end{figure*}
%%%%

We begin by considering the behavior of the $\Nsubnobeta{2,1}$ and $D_2$ observables in the limit of two resolved subjets. 
In terms of the observables, the resolved region of phase space is defined by the conditions
%%%%
\begin{align}\label{eq:resolved_limit}
\Nsub{2,1}{\beta}\ll 1\,, \qquad \text{or} \qquad  D_2^{(\beta)}\ll 1\,.
\end{align}
%%%%
In this regime, the jet is comprised of two subjets, which are either collinear with similar energy fractions, or have a wide opening angle and hierarchical energy fractions. We will refer to these as the collinear subjets and soft subjet configurations, respectively. They are shown schematically in \Fig{fig:jet_configs}. These configurations have been discussed in detail in \Refs{Bauer:2011uc,Larkoski:2015zka,Larkoski:2015kga}, and we therefore do not elaborate further.

An important simplification which occurs in the resolved limit is a natural factorization into a process which produces the jet substructure, and the emission of soft and collinear radiation from the hard substructure, which occurs at longer time scales. This factorization can be made rigorous, for example, within the soft-collinear effective theory (SCET) \cite{Bauer:2000yr,Bauer:2001ct,Bauer:2001yt,Bauer:2002nz}. Because of the additional hierarchies associated with the substructure of the jet, additional modes are also typically required \cite{Bauer:2011uc,Larkoski:2015kga,Larkoski:2015zka,Neill:2015nya,piotr_talk,piotr_new}. We shall refer to the class of effective field theories required to study this region of phase space as SCET$_+$. In particular, effective field theory descriptions exist for both the collinear subjets  \cite{Bauer:2011uc} and soft subjet configurations \cite{Larkoski:2015kga,Larkoski:2015zka}, allowing for an all orders treatment of jet substructure observables measured in this limit. A method for combining the effective field theory descriptions of the soft subjet and collinear subjets regions of phase space was given in \Ref{Larkoski:2015kga}, and studied for the particular case of the $D_2$ observable.

In the resolved limit, there is also a considerable simplification for the particular case of the  $\Nsubnobeta{2,1}$ observable. At leading power, for $\beta>1$, the different $N$-subjettiness axes definitions are equivalent. In particular, the two axes defining $\Nsub{2}{\beta}$ will align with the subjets, and any differences in the exact position of the axes are power corrections. This is well known from the study of exclusive $N$-jettiness \cite{Stewart:2010tn}. For the case of $N$-subjettiness, this leads to a universal behavior in the resolved limit. The assumption of $\beta>1$ is important so that recoil effects \cite{Catani:1992jc,Dokshitzer:1998kz,Banfi:2004yd,Larkoski:2013eya} are power suppressed. For $\beta\leq 1$, recoil effects are leading power, reintroducing dependence on the axes definition. This is familiar from the event shape broadening \cite{Catani:1992jc,Dokshitzer:1998kz,Becher:2011pf,Chiu:2012ir,Becher:2012qc}. Universality can be restored through the use of recoil-free axes \cite{Banfi:2004yd,Larkoski:2014uqa}, which are insensitive to soft recoil. Recoil-free axes include both minimized axes as well as those defined with WTA recombination. In this section we restrict our attention to axes defined with WTA recombination, leaving a discussion of minimization to \Sec{sec:physical_region}.

%%%%
\begin{figure*}[t]
\centering
\subfloat[]{\label{fig:sing_close}
\includegraphics[width=6.5cm]{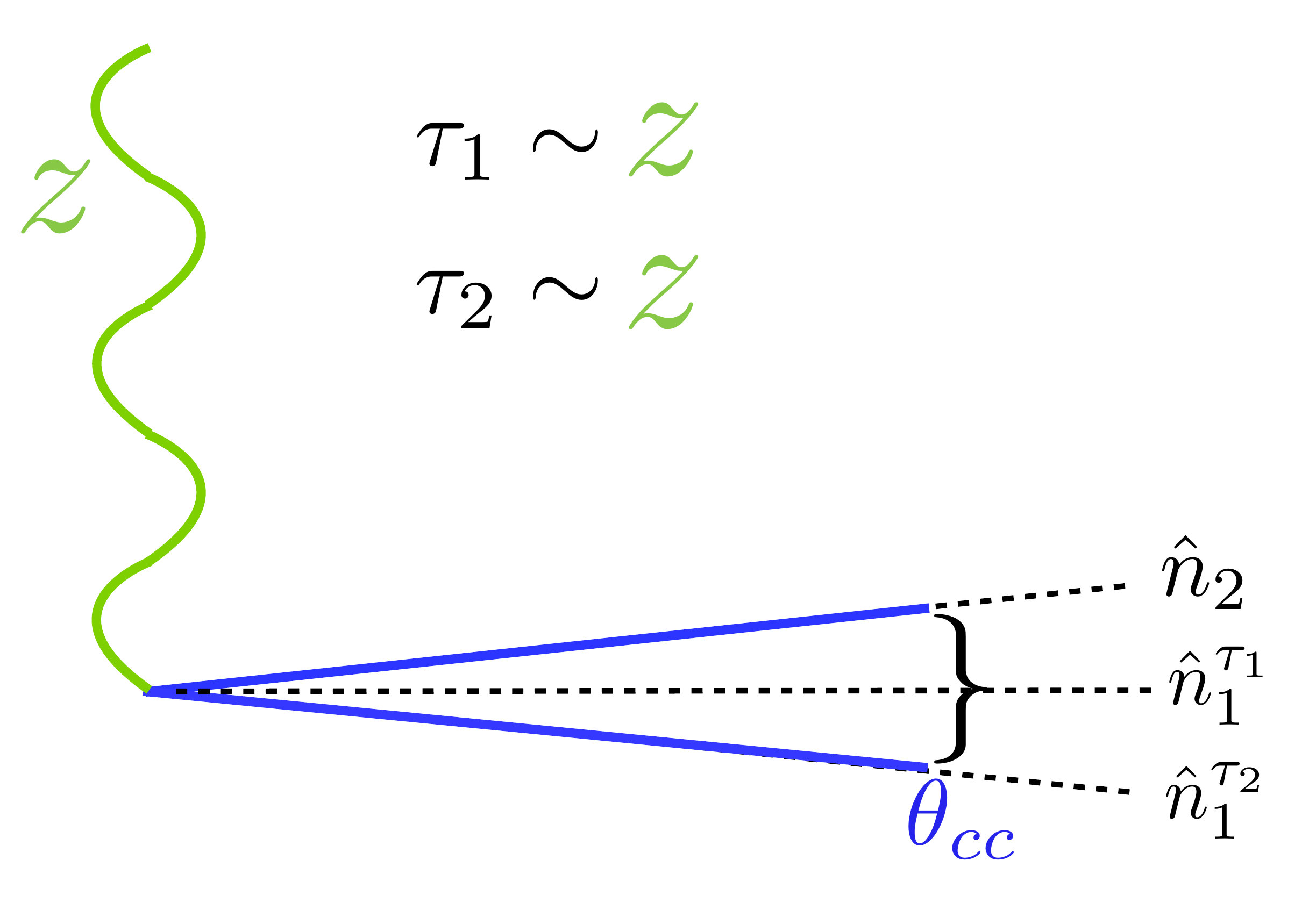} 
}\hspace{1cm}
\subfloat[]{\label{fig:sing_wide}
\includegraphics[width=7.5cm]{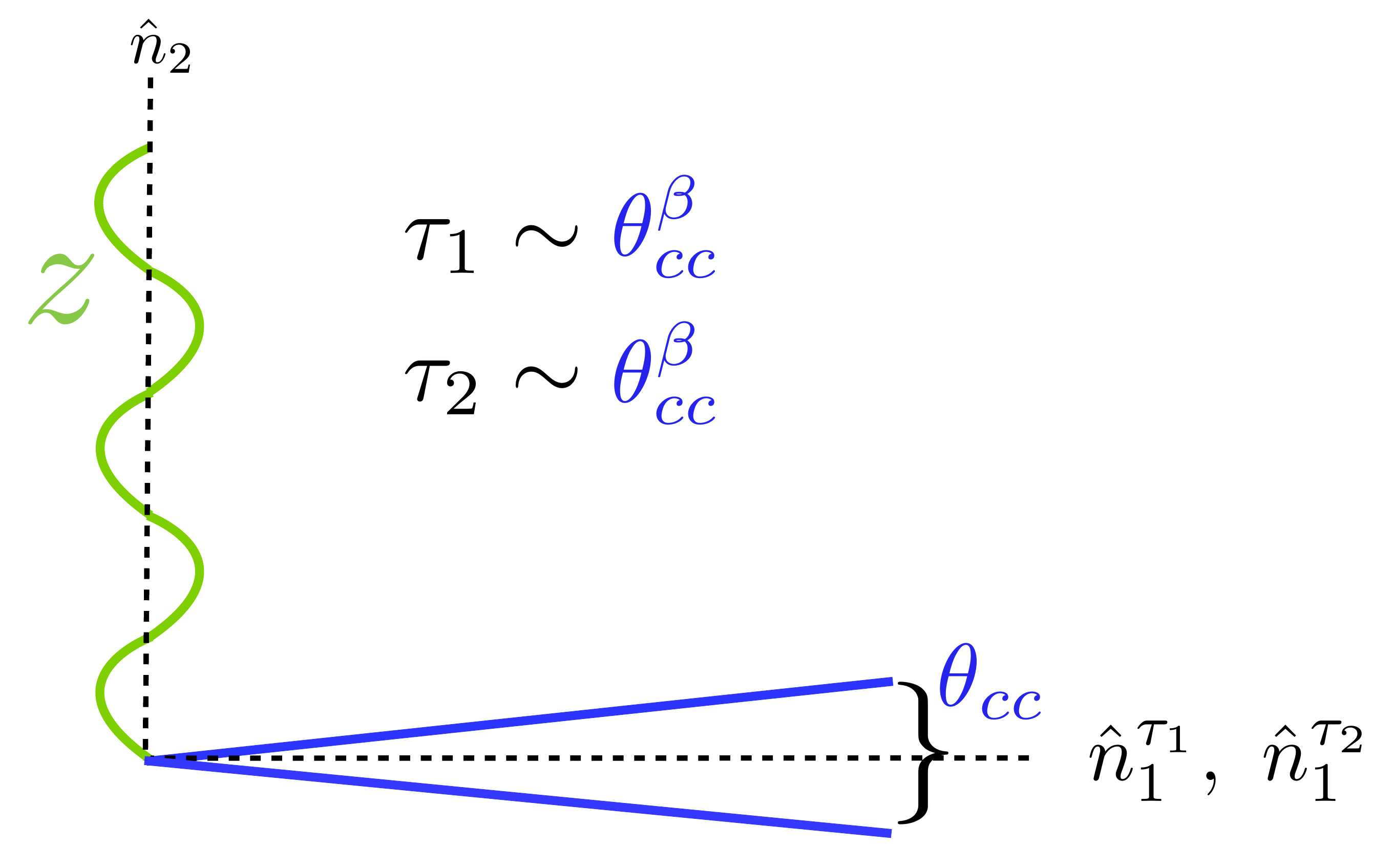}    
} 
\caption{The singular configurations involving three partons within a jet which are relevant for studying the  $\Nsubnobeta{2,1}\to 1$ limit. In configuration (a), both observables are set by a soft emission, giving rise to an unregulated collinear singularity. In configuration (b), both observables are set by a collinear splitting, giving rise to an unregulated soft singularity.
}
\label{fig:singular_configs}
\end{figure*}
%%%%

To demonstrate explicitly the universality in the resolved limit, we calculate the $\Nsubnobeta{2,1}$ observable with different axes definitions in fixed order perturbation theory and in parton shower Monte Carlo. We consider $e^+e^-$ collisions at a center of mass energy of $1$ TeV, and select the hardest hemisphere jet. We place a mass cut of $[80,100]$ GeV to emulate the situation as relevant for boosted $Z$ boson identification. Events are clustered and analyzed with \fastjet{3.1.3} \cite{Cacciari:2011ma,fjcontrib}. When necessary, we have modified the implementations in \fastjet{}, such that all axes, measures, and recombination schemes are those appropriate for $e^+e^-$ collisions.

Fixed order events are generated with \nlojet{4.1.3} \cite{Nagy:1997yn,Nagy:1998bb,Nagy:2001xb,Nagy:2001fj,Nagy:2003tz}. The observables that we are considering, namely $\Nsubnobeta{2,1}$ and $D_2$, are first non-zero in perturbation theory when there are three partons within a jet. We will refer to the leading order (LO) distribution as that generated from tree level matrix elements for $e^+e^-\to 4$ partons, which are $\mathcal{O}(\alpha_s^2)$. We will also study the behavior at next to leading order (NLO), which is calculated using the 1-loop matrix elements for $e^+e^-\to4$ partons, and tree level matrix elements for $e^+e^-\to 5$ partons. Perturbative scale variations obtained by varying the renormalization scale by a factor of two will be shown for all fixed order predictions. Particularly at NLO, there is also a non-negligible statistical uncertainty due to the fact that our selection cuts are highly inefficient. Statistical uncertainties are not shown, but are clear from the jitter in the distributions. They do not affect our classification of the singular structures.

Parton shower events are generated using \vincia{1.2.02} \cite{Giele:2007di,Giele:2011cb,GehrmannDeRidder:2011dm,Ritzmann:2012ca,Hartgring:2013jma,Larkoski:2013yi}. We have chosen to use \vincia{} as it has been demonstrated to agree well with analytic calculations of two-prong substructure observables \cite{Larkoski:2015kga}.

In \Fig{fig:nsub_FO_2} we compare $\Nsubnobeta{2,1}$ distributions with $\beta=1$ and $\beta=2$. Three different definitions of the $N$-subjettiness axes, namely  exclusive $k_T$ axes, exclusive C/A axes and generalized $k_T$ axes with $p=1/2$, which we refer to as $k_T^{1/2}$, are considered. In the case of $\beta=1$, a WTA recombination scheme has been used to eliminate recoil.  Fixed order results at LO are shown in the left column and parton level \vincia{} distributions are shown in the right column, both in a logarithmic scale. In the resolved limit, $\Nsubnobeta{2,1}\to 0$, the observable as defined with different axes definitions exhibits a universal behavior, both in fixed order perturbation theory, and in the parton shower result. This should be contrasted with the behavior in the unresolved limit, where for the three different axes choices, all the different behaviors of \Fig{fig:sing_structure} are obtained, namely an endpoint divergence, a smoothly vanishing distribution, and a shoulder in the physical region. It is interesting to note that the XCone recommended seed axes are $k_T^{1/2}$ for $\beta=2$ and WTA $k_T$ for $\beta=1$. For both $\beta=1$ and $\beta=2$ these lead to a shoulder in the physical region.

%%%%
\section{Generalized $k_T$ Algorithms and the $\Nsubnobeta{2,1}\to 1$ Endpoint}\label{sec:endpoint}
%%%%

%%%%%%%%
\begin{figure*}[t]
\centering
\subfloat[]{\label{fig:wkt1}
\includegraphics[width=7.5cm]{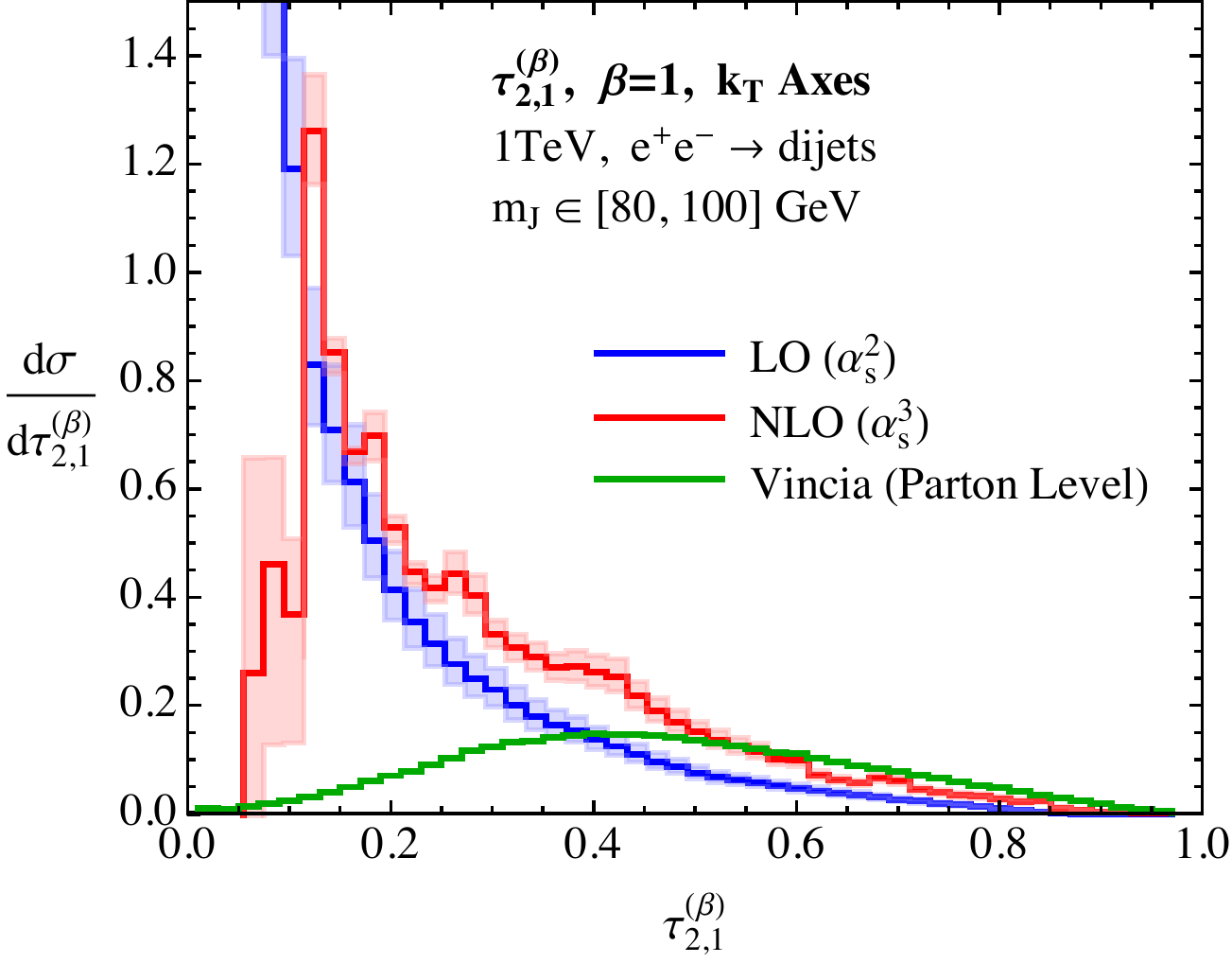}    
} \hspace{1cm}
\subfloat[]{\label{fig:wkt2}
\includegraphics[width=7.5cm]{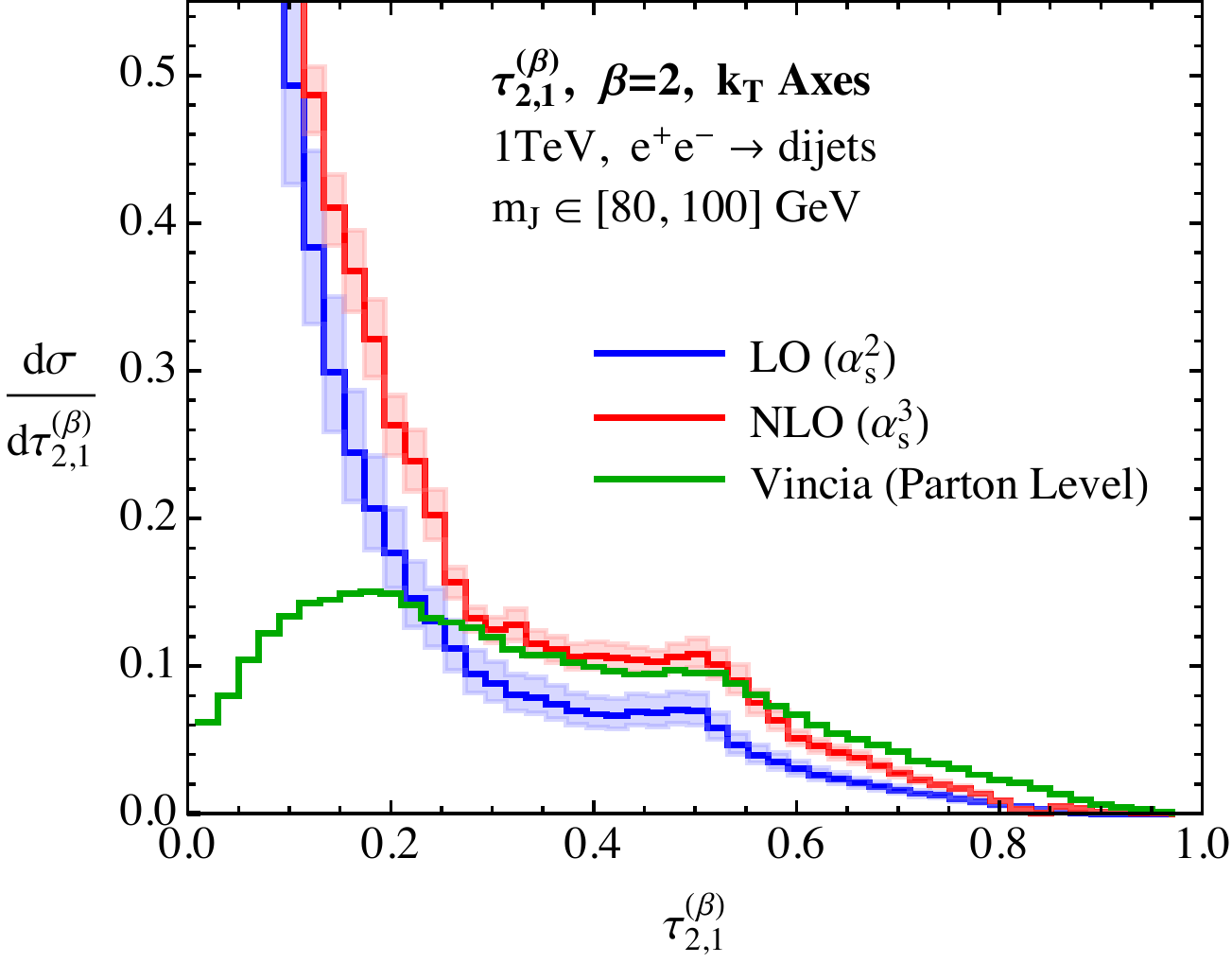} 
}
\caption{The behavior of the $\Nsubnobeta{2,1}$ observable with axes defined by the exclusive $k_T$ algorithm for $\beta=1$ in (a), and $\beta=2$ in (b). Fixed order results are shown at LO and NLO, and contrasted with parton level \vincia{} Monte Carlo. }
\label{fig:wkt}
\end{figure*}
%%%%%%%%

From the definition of the $N$-subjettiness observables in \Eq{eq:Nsub_def}, it is clear that the ratio observable $\Nsubnobeta{2,1}$ exhibits a physical endpoint at $\Nsubnobeta{2,1}=1$. This is a natural place to expect interesting behavior. In this section we will analyze the behavior as $\Nsubnobeta{2,1}\to1$ for $N$-subjettiness axes  defined using the generalized $k_T$ metric of \Eq{eq:gen_kt_metric} with $p\geq 0$.

To understand the $\Nsubnobeta{2,1}\to 1$ endpoint behavior we must consider singular configurations of soft and collinear particles, which can contribute at this endpoint. We will begin by analyzing such configurations when there are three particles within the jet (LO), but the analysis will straightforwardly generalize. Since we are interested in the behavior as $\Nsubnobeta{2,1}\to 1$, in the most singular region of the phase space both $\Nsubnobeta{1}$ and $\Nsubnobeta{2}$ must be set to leading power by the same emission. The two singular configurations relevant for the $\Nsubnobeta{2,1}\to 1$ endpoint behavior are shown in  \Fig{fig:singular_configs}. In the configuration of \Fig{fig:sing_close} both axes lie in the collinear sector, where the collinear particles are characterized as having ${\cal O}(1)$ energy fraction and an angular separation of $\theta_{cc}$. The values of both $\Nsubnobeta{1}$ and $\Nsubnobeta{2}$ are set by the wide angle soft emission. The singular configuration occurs when the angle between the two collinear particles goes to 0. In the configuration of \Fig{fig:sing_wide} one axis lies on a wide angle soft particle, with characteristic energy $z_s$. The values of both $\Nsubnobeta{1}$ and $\Nsubnobeta{2}$ are set by the collinear emission. The singularity then occurs as $z_s\to0$.

 The endpoint behavior for the generalized $k_T$ metric given in \Eq{eq:gen_kt_metric} can now be easily understood by considering the compatibility of the clustering metric with the different singular configurations. In particular, one must understand for what values of $p$ in the generalized $k_T$ metric the axes will be configured as in \Fig{fig:sing_close} or \Fig{fig:sing_wide}. We will denote the generalized $k_T$ metric distance of \Eq{eq:gen_kt_metric} as measured between two soft particles, by $d_{ss}$, as measured between a soft and collinear particle, by $d_{sc}$, and as measured between two collinear particles, by $d_{cc}$.

 We will first show that with a mass cut, the configuration in \Fig{fig:sing_close} does not give rise to an endpoint singularity for any value of $p$. For both axes to lie in the collinear sector, the collinear particles must be clustered last. Therefore, we must have
 \begin{align}\label{eq:close_metric}
 d_{ss} \sim d_{sc}\sim z^{2p} \ll d_{cc} \sim \theta_{cc}^2\,.
 \end{align}
 For $p=0$, this condition is already inconsistent, so we focus on  $p>0$. For the configuration shown in  \Fig{fig:sing_close} to have a collinear singularity, we must have the relation
 \begin{align}\label{eq:close_pc}
 z> \theta_{cc}^\beta\,.
 \end{align}
Recall that $\beta>0$ for IRC safety. In particular, with a mass cut, the soft wide angle emission sets the mass, so 
 \begin{align}\label{eq:close_mass}
 z\sim \frac{m^2}{E_J^2}\,.
 \end{align}
 Depending on the value of $\beta$, the relations in \Eqs{eq:close_metric}{eq:close_pc} are either inconsistent, or if they are consistent, the angle $\theta_{cc}$ is regulated by the mass cut. Therefore for all values of $p$ with the generalized $k_T$ metric, this configuration does not give rise to a singularity at the $\Nsubnobeta{2,1}\to 1$ endpoint.

 We now consider the configuration in \Fig{fig:sing_wide}. The clustering constraint that the axis lies on the wide angle soft emission is
  \begin{align}\label{eq:wide_metric}
 d_{ss} \sim d_{sc}\sim z^{2p} \gg d_{cc} \sim \theta_{cc}^2\,,
 \end{align}
 and the consistency relation that both $\Nsubnobeta{1}$ and $\Nsubnobeta{2}$ are set by the collinear emission is
 \begin{align}\label{eq:wide_pc}
 z< \theta_{cc}^\beta\,,
 \end{align}
 and
  \begin{align}\label{eq:close_mass1}
\theta_{cc} \sim \frac{m}{E_J}\,.
 \end{align}
 
 For $p>0$, depending on the value of $\beta$, the relations of \Eq{eq:wide_metric} and \Eq{eq:wide_pc} are either inconsistent, or $z$ is regulated by the mass cut. Therefore, this configuration does not contribute to a singularity at the $\Nsubnobeta{2,1}\to 1$ endpoint.
 For example, for the case of exclusive $k_T$ axes, the mass cut implements a minimum angular separation between the two $N$-subjettiness axes, which we denote $\theta_{n_1, n_2}$. From \Eqs{eq:close_metric}{eq:close_mass}, which describe the configuration in \Fig{fig:sing_close} where the $2$-subjettiness axes are close, we find
 %%%%
 \begin{align}\label{eq:axes_sep}
 \theta_{n_1 n_2} \gtrsim \frac{m_J^2}{E_J^2}\,.
 \end{align}
 %%%%
Since we will always implement a mass cut of $[80,100]$ GeV for $500$ GeV jets, the possible singular behavior resulting from \Fig{fig:sing_close} is well regulated, and we expect a fixed order perturbative description to be applicable.

 However, for $p=0$ the relations in \Eqs{eq:wide_metric}{eq:wide_pc} are consistent, and the wide angle soft particle can become arbitrarily soft, remaining consistent with both the mass cut and the clustering conditions, giving rise to a singularity at the $\Nsubnobeta{2,1}\to 1$ endpoint. Although we have presented the argument for the case of $3$ particles, it extends  to an arbitrary number of soft and collinear emissions at the endpoint.
 
 %%%%%%%% 
 \begin{figure*}[t]
\centering
\subfloat[]{\label{fig:CA1}
\includegraphics[width=7.5cm]{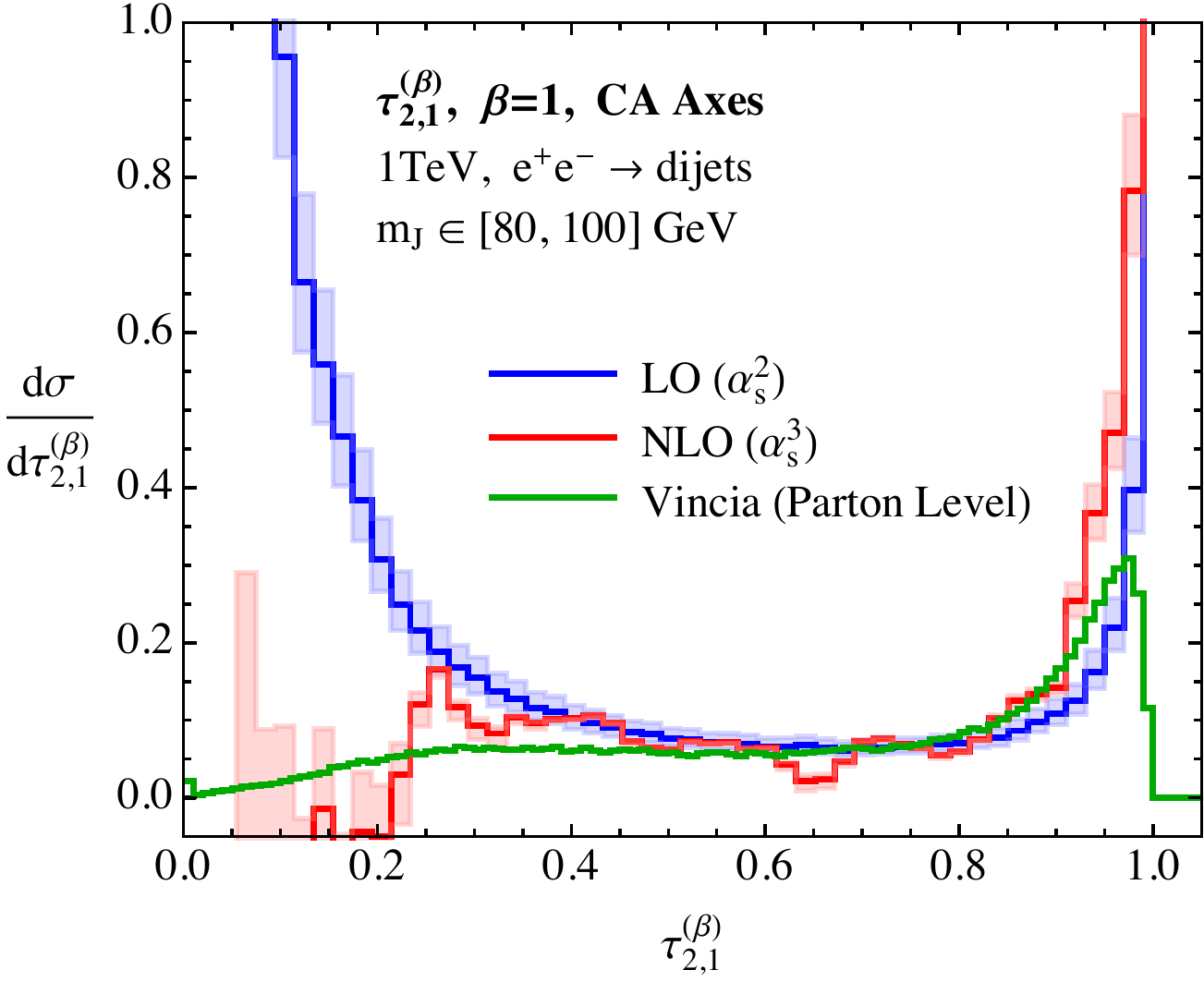}    
} \ \ \hspace{1cm}
\subfloat[]{\label{fig:CA2}
\includegraphics[width=7.5cm]{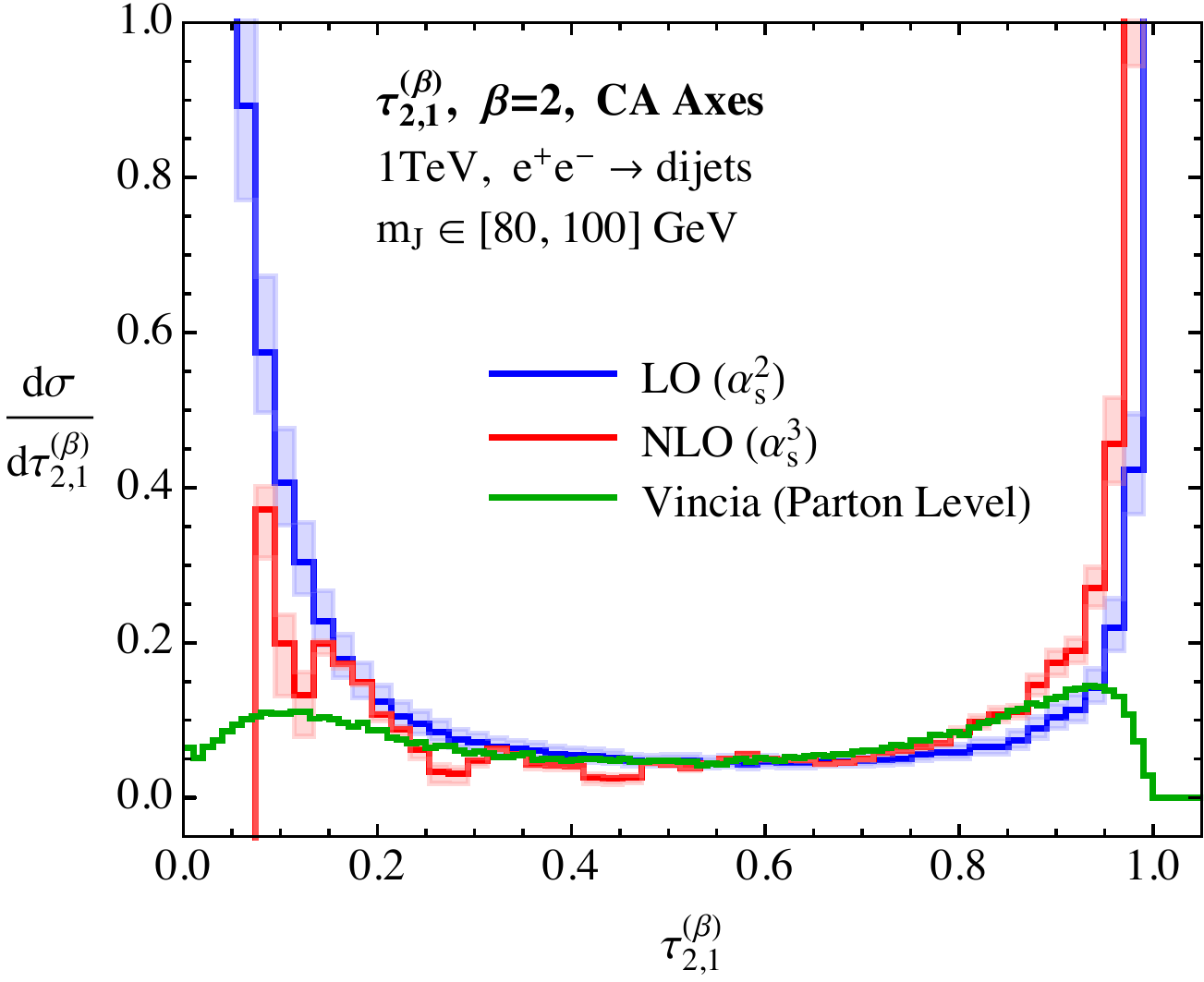} 
}
\caption{The behavior of the $\Nsubnobeta{2,1}$ observable with axes defined by the exclusive C/A algorithm for $\beta=1$ in (a), and $\beta=2$ in (b). Fixed order results are shown at LO and NLO, and contrasted with parton level \vincia{} Monte Carlo. 
}
\label{fig:CA}
\end{figure*}
%%%%%%%% 

In summary, we have shown that  for $p>0$, no singularity exists at the $\Nsubnobeta{2,1}\to 1$ endpoint, while for $p=0$ there is a soft singularity at the endpoint. This dependence on the axes definitions arises because the singular behavior at the endpoint is being regulated by the ability of the axes definitions to avoid the configurations in  \Fig{fig:singular_configs}. Computationally, this is unsatisfactory, as it implies that the axes must ``flop" so as to avoid these singular regions.

Having isolated the particular axes configuration that leads to the divergence for $p=0$, namely that of \Fig{fig:sing_wide}, it is straightforward to write down a factorization theorem describing the behavior in this singular region of phase space, for example, using effective field theory techniques. However, because the non-trivial behavior described by such a factorization theorem begins at higher orders in perturbation theory, such a study is beyond the scope of this paper. Instead, to demonstrate these two types of possible endpoint behaviors, we will consider as case studies, the $N$-subjettiness axes as defined with the C/A $(p=0)$, and $k_T$ $(p=1)$ axes. We will compute both  these cases in fixed order perturbation theory and parton shower Monte Carlo. The event selection is the same as detailed in \Sec{sec:resolved}.

In \Fig{fig:wkt} we show results for $\Nsubnobeta{2,1}$ as defined with the exclusive $k_T$ axes, and in \Fig{fig:CA} with the exclusive C/A algorithm. The LO and NLO fixed order distributions as well as parton level results from \vincia{} are shown. For both $k_T$ and C/A axes, the divergence as $\Nsubnobeta{2,1}\to 0$ is present and the parton shower implements the expected Sudakov resummation. On the other hand, while a divergence at LO and NLO is clearly observable in \Fig{fig:CA} when $\Nsubnobeta{2,1}$ is defined with C/A axes, no such divergence is observed with $k_T$ axes, and the $\Nsubnobeta{2,1}\to1$ behavior is well described by fixed order perturbation theory.

%%%%%%%%%%%
\begin{figure*}[t]
\centering
\subfloat[]{\label{fig:endpoint_a}
\includegraphics[width=8.5cm]{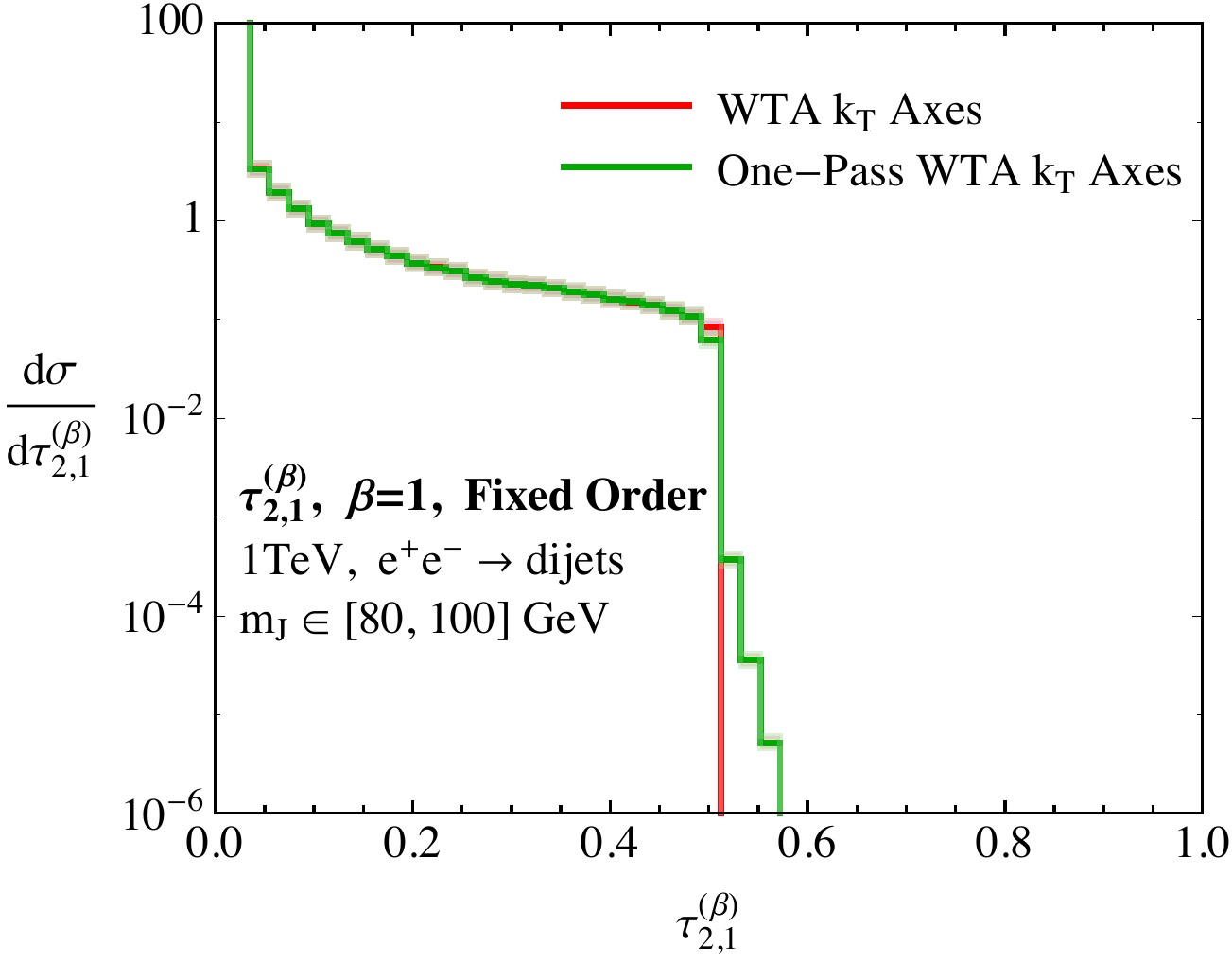}    
} \hspace{0.2cm}
\subfloat[]{\label{fig:endpoint_b}
\includegraphics[width=8.5cm]{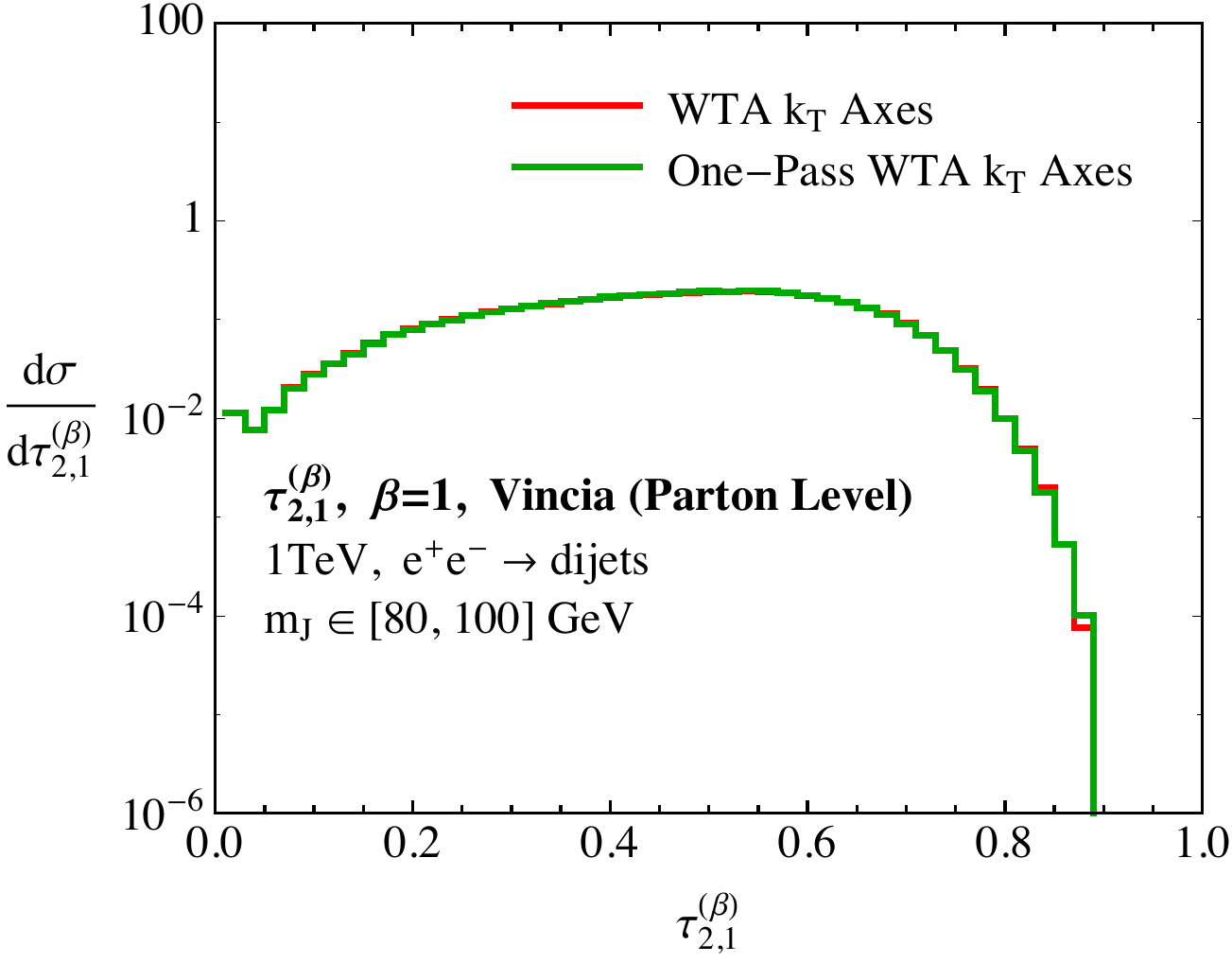} 
}
\caption{A comparison of $\Nsubnobeta{2,1}$ with $\beta=1$ for WTA $k_T$ axes and one-pass minimized WTA $k_T$ axes in leading order perturbation theory in (a) and \vincia{} parton level Monte Carlo in (b). A clear shoulder at $\Nsubnobeta{2,1}=0.5$ is observed in fixed order perturbation theory. It is extended slightly by minimization, as described in the text.
}
\label{fig:endpoint}
\end{figure*}
%%%%%%%%%%%

%%%%
\section{Minimization, Winner Take All Recombination, and Singularities in the Physical Region}\label{sec:physical_region}
%%%%

In this section we turn to the study of the $\Nsubnobeta{2,1}$ observable for the phenomenologically most relevant axes choices, namely minimized axes, and axes defined with the Winner Take All (WTA) recombination scheme. These are the definitions which are currently used by the ATLAS and CMS experiments, and which achieve the best discrimination power. We will demonstrate that with these choices of axes, the $\Nsubnobeta{2,1}$ observable exhibits discontinuities in the physical region, which lead to regulated singularities at higher orders in perturbation theory.  This behavior has severe consequences for the analytic tractability of these observables, as high particle multiplicities are required to fill out the physical region. This implies that there is little overlap between $\Nsubnobeta{2,1}$ calculated at low orders in perturbation theory, and from parton shower Monte Carlo simulations. This behavior is expected for certain jet substructure observables, for example, Q-jet \cite{Ellis:2012sn} variables where $\mathcal{O}(10)$ particles are required to get non-trivial distributions. However, it is more surprising for the case of $N$-subjettiness, which is similar to traditional jet shape observables. In \Sec{sec:non_pert}, we will show that this complicated singular structure also leads to large non-perturbative effects.

Before continuing to discuss the particular case of $\Nsubnobeta{2,1}$ it is worth briefly reviewing the possibility of singularities in the physical region  \cite{Catani:1997xc}. Interestingly, these can occur even for observables which satisfy traditional definitions of IRC safety \cite{Sterman:1977wj,Ellis:1991qj}.  Such singularities arise from discontinuities within the physical region at a given order in perturbation theory. Discontinuities can arise due to a physical constraint on the observable arising from having a fixed number of partons in the final state. If this constraint is no longer satisfied when additional partons are added at higher orders in perturbation theory, then there will in general be a divergence at the location of the discontinuity, due to a miscancellation of real and virtual corrections. Resummation is then required to achieve an accurate perturbative prediction for the cross section. Singularities in the physical region are familiar from classic $e^+e^-$ event shapes such as the $C$-parameter \cite{Parisi:1978eg,Donoghue:1979vi}, which exhibits a singularity at $C=3/4$ due to a discontinuity at this location in the $\mathcal{O}(\alpha_s)$ distribution \cite{Ellis:1980wv}, and the Fox-Wolfram moments \cite{Fox:1978vw,Fox:1978vu,Fox:1979id}.\footnote{The second Fox-Wolfram moment, $H_2$, and the $C$-parameter are directly related for massless particles, so that these singularities have the same origin.} The resummation of singular contributions at $C=3/4$ was performed in \Ref{Catani:1997xc}.

Minimized axes for  $N$-subjettiness are defined as those axes $\hat n_i$ which minimize the sum in \Eq{eq:Nsub_def}. For practical purposes, this minimization is performed starting from seed axes, which are typically found using  an exclusive jet clustering algorithm.  As shown in \Ref{Stewart:2015waa}, and implemented in the XCone algorithm, the seed axes can be optimized by matching the clustering metric to the $N$-jettiness measure for an arbitrary $\beta$. We will study the cases $\beta=1$ and $\beta=2$ in this section. The recommended XCone seed axes for these cases are WTA $k_T$ axes for $\beta=1$, and generalized $k_T$, with $p=1/2$ and $E$-scheme recombination for $\beta=2$. Because of this close link between WTA and minimized axes for $\beta=1$, we have decided to consider them together in this section. However, it is important to note that they are not identical. In particular, with WTA recombination, the axes lie on particles within the jet. Since we will always choose to match the jet clustering procedure with the $N$-subjettiness measure we can view the WTA versions as minimization with the constraint that all the jet axes must lie on a particle within the jet. This is not in general true for the exact minimized axes. For example, if we consider three particles with equal energy and take $\beta=1$, if there is no angle greater than $120^\circ$, then the minimized axis lies at the Fermat point inside the triangle.  (The Fermat point is the point at which the angles between all lines extending to the vertices of the triangle is $120^\circ$.)  This exact minimization typically has a larger effect on $\Nsubnobeta{1}$, leading to larger values of $\Nsubnobeta{2,1}$ when one-pass minimization is applied.

%%%%
\begin{figure*}[t]
\centering
\subfloat[]{\label{fig:compare_min_a_2}
\includegraphics[width=8.5cm]{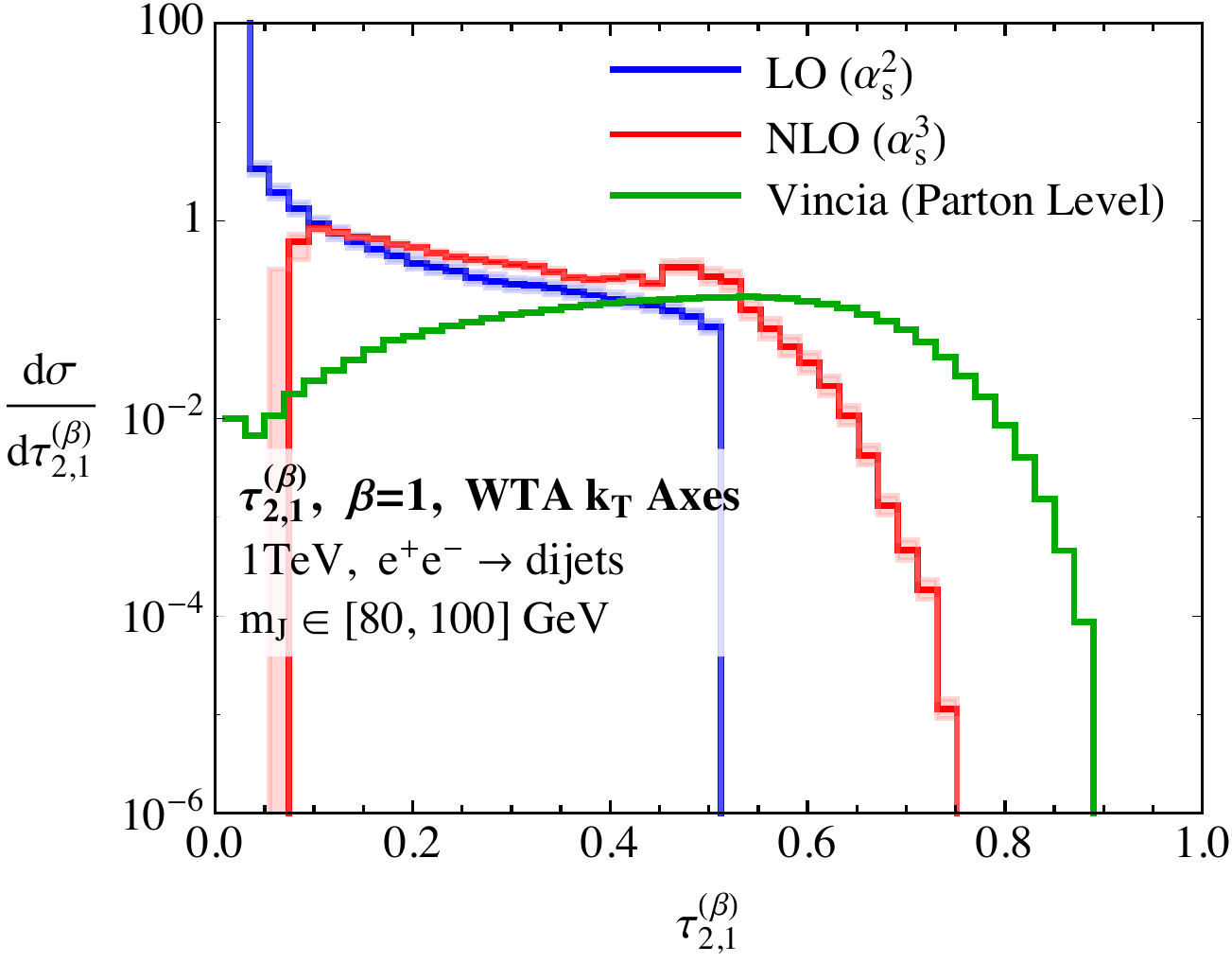}    
} \ \ 
\subfloat[]{\label{fig:compare_min_b_2}
\includegraphics[width=8.4cm]{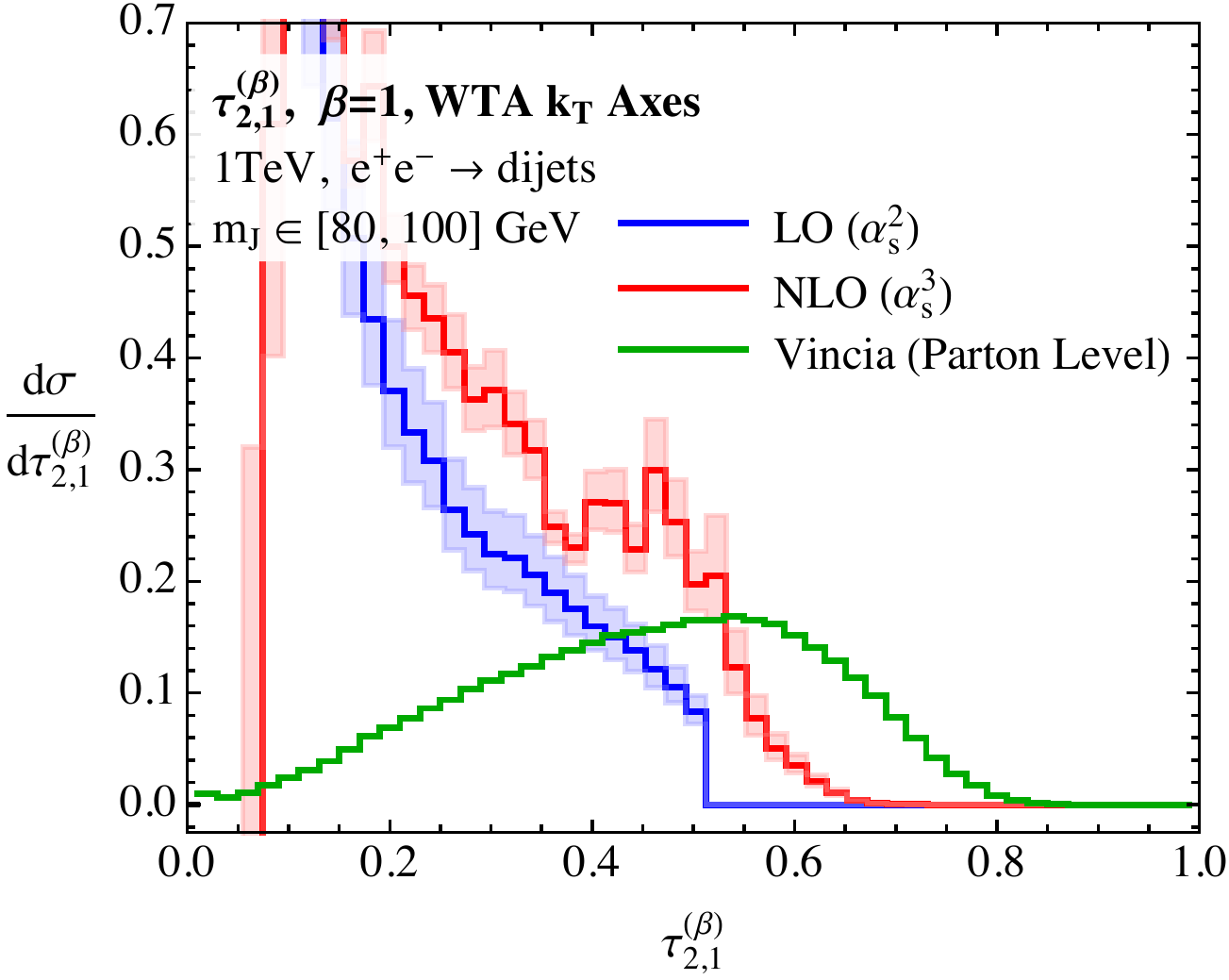} 
}
\caption{Distributions for $\Nsubnobeta{2,1}$ as defined using WTA $k_T$ axes with $\beta=1$ in both fixed order perturbation theory and parton level Monte Carlo. The distributions are shown on a logarithmic scale in (a), and on a linear scale in (b). The behavior observed in Monte Carlo as $\Nsubnobeta{2,1}\to1$ is not well reproduced in fixed order perturbation theory.
}
\label{fig:wtavop2}
\end{figure*}
%%%%

We begin by considering the case of $\beta=1$ with WTA $k_T$ and one-pass WTA $k_T$ axes. With $N$-particles in the final state, with WTA axes, for $\Nsubnobeta{1}$, $N-1$ particles contribute, while for $\Nsubnobeta{2}$, $N-2$ particles contribute. The maximum value of $\Nsubnobeta{2,1}$ is therefore $\Nsubnobeta{2,1}=(N-2)/(N-1)$. For $N=3$, this gives $\Nsubnobeta{2,1}=1/2$, and for $N=4$, this gives $\Nsubnobeta{2,1}=2/3$. Interestingly, both of these values are well within the physical region for $\Nsubnobeta{2,1}$, and furthermore, the volume of phase space contributing at the endpoint is in general non-zero. We therefore expect that shoulders (discontinuities) will appear in the physical region for these values, and corresponding singularities at higher orders in perturbation theory. By considering configurations of three particles, it can be shown that exact minimization extends the endpoint to $1/\sqrt{3}\simeq 0.577$.

To verify this behavior, we now perform a study of minimized and WTA axes  in fixed order perturbation theory and parton shower Monte Carlo.  In \Fig{fig:endpoint_a} we show the behavior of $\Nsubnobeta{2,1}$ as defined with WTA $k_T$ axes, and with one-pass minimization starting from WTA $k_T$ axes. The WTA $k_T$ axes are the XCone recommended seed axes, and therefore we expect the one-pass minimization to have a relatively minor effect. However, we see that it does extend the physical region at LO from $\Nsubnobeta{2,1}=0.5$ to $\Nsubnobeta{2,1}\simeq0.577$, as expected. The fraction of events which extend to the truly minimized endpoint after one-pass minimization is quite small, showing that the local minima are deep.

Since the restriction on the maximal value of $\Nsubnobeta{2,1}$ arises due to the low number of particles in the final state, in \Fig{fig:endpoint_b} we show the same distributions in \vincia{} Monte Carlo at parton level. The physical region is greatly extended by the greater multiplicity of particles in the final state, and remarkably, its peak is beyond the support of the LO distribution. The endpoint differences due to the application of the one-pass minimization are also removed.

Due to the presence of the discontinuity in the physical region, it is particularly interesting to consider the behavior of the NLO fixed order corrections. In \Fig{fig:wtavop2}  we consider the NLO corrections to $\Nsubnobeta{2,1}$ as defined with WTA $k_T$ axes,\footnote{Similar behavior is observed for  the one-pass minimized axes.} and compare with the leading order perturbative result and the result from parton level \vincia{}. The NLO contribution extends the distribution beyond $\Nsubnobeta{2,1}=0.5$, up to the expected endpoint at $\Nsubnobeta{2,1}=2/3$. In the log plot we can see that a few events go beyond this point, but it is quite well respected. However, this is still not sufficient to begin to describe the shape of the Monte Carlo distribution in the unresolved region, suggesting that even higher orders in perturbation theory are required.  The NLO contributions also give a sharp peak around $\Nsubnobeta{2,1}=0.5$ demonstrating the lack of cancellation between real and virtual corrections. 

%%%%%%%%%%%
\begin{figure*}[t]
\centering
\subfloat[]{\label{fig:endpoint_a_beta2}
\includegraphics[width=8.5cm]{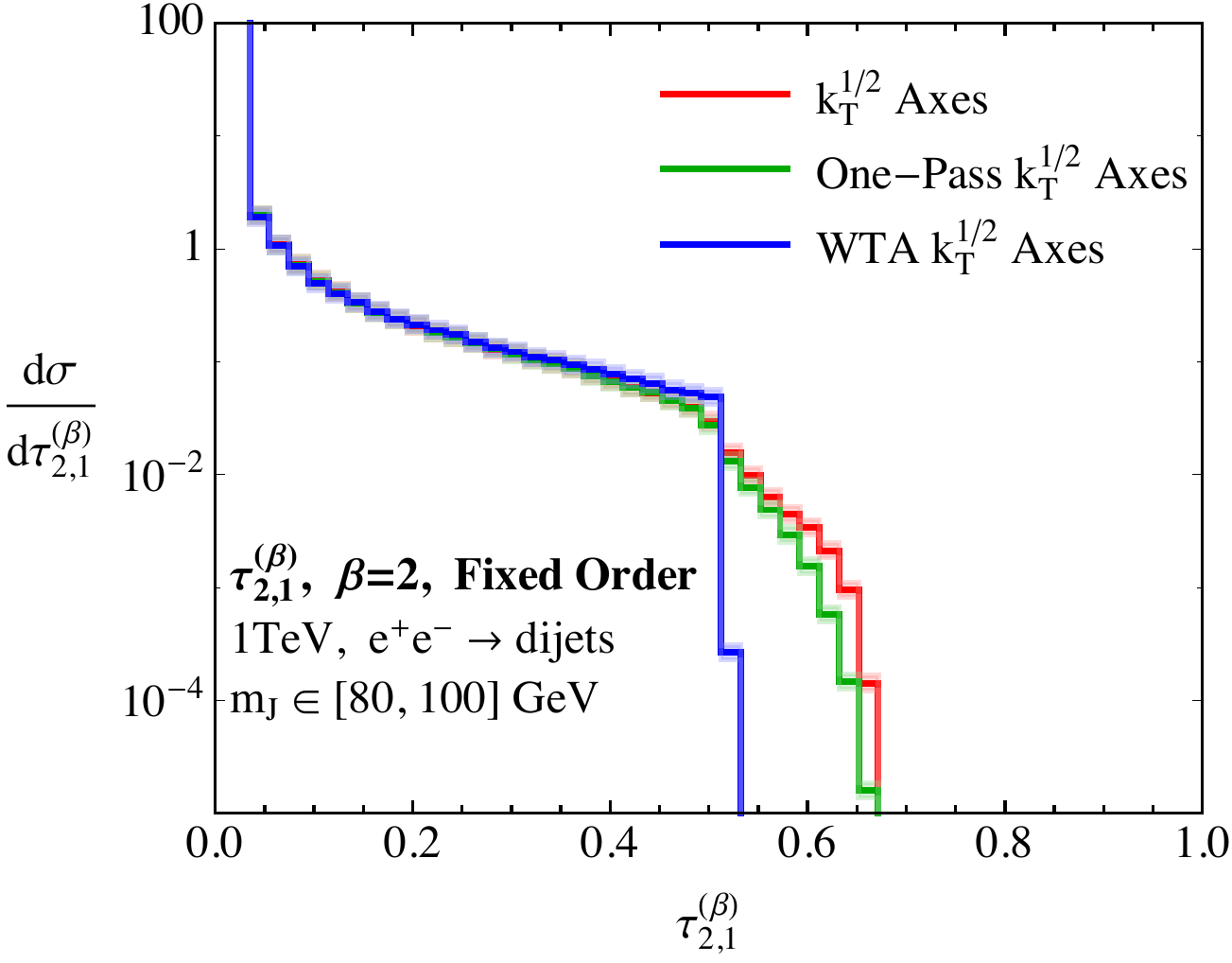}    
} \ \ \hspace{0.2cm}
\subfloat[]{\label{fig:endpoint_b_beta2}
\includegraphics[width=8.5cm]{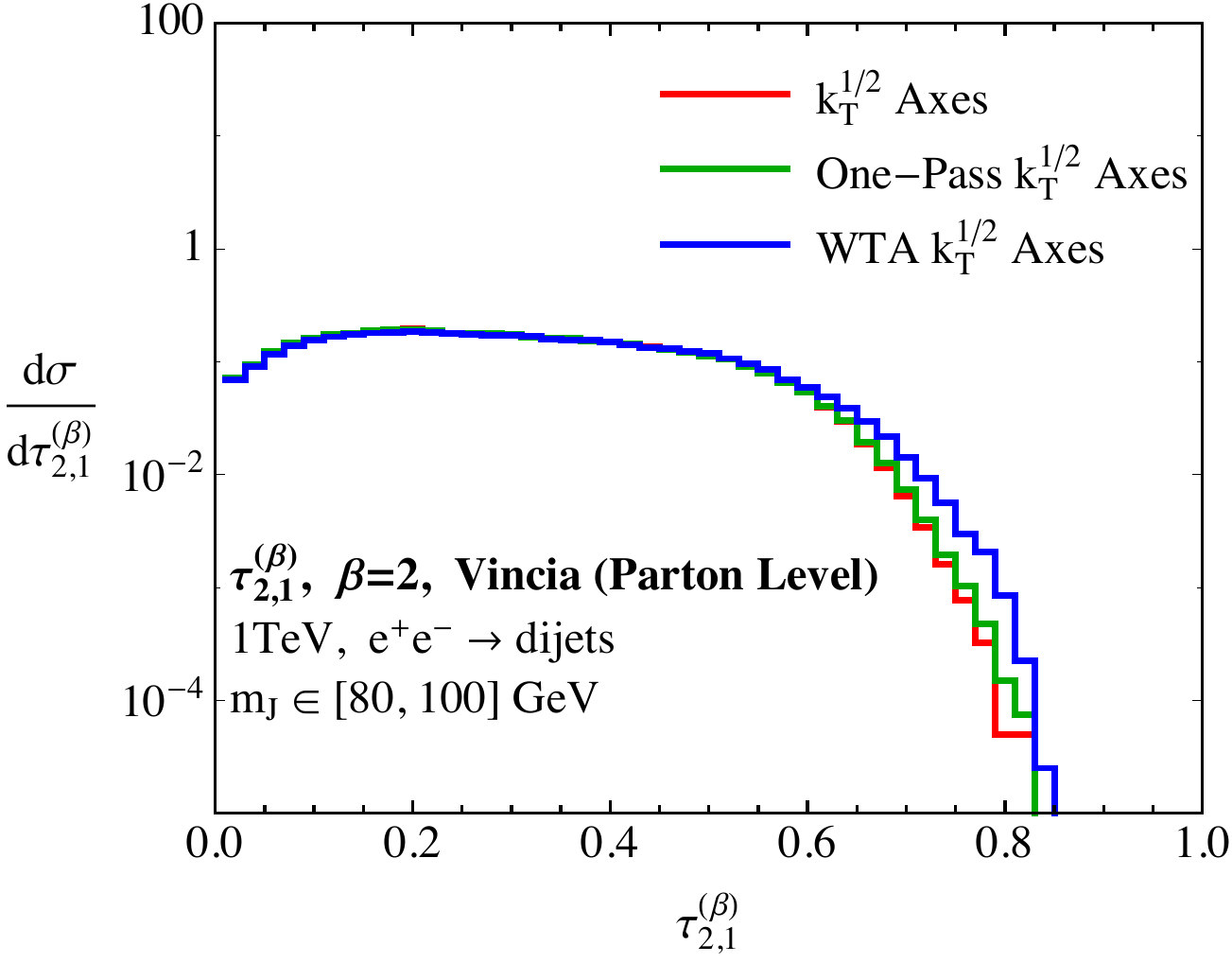} 
}
\caption{A comparison of $\Nsubnobeta{2,1}$ with $\beta=2$ for $k_T^{1/2}$, one-pass minimized $k_T^{1/2}$ and WTA $k_T^{1/2}$ axes in leading order perturbation theory in (a) and \vincia{} parton level Monte Carlo in (b).
}
\label{fig:endpoint_beta2}
\end{figure*}
%%%%%%%%%%%

Having understood the behavior of minimized and WTA axes for $\beta=1$, we now show that a similar behavior persists for $\beta=2$. For $\beta=2$, the XCone recommended seed axes which match the clustering with the $N$-subjettiness minimization are generalized $k_T$ with $p=1/2$ and $E$-scheme recombination, which we refer to as $k_T^{1/2}$.

In \Fig{fig:endpoint_a_beta2} we compare the behavior of $\Nsubnobeta{2,1}$ as defined with $k_T^{1/2}$ axes, one-pass minimized $k_T^{1/2}$ axes, as well as WTA $k_T^{1/2}$ axes to understand how the distribution is modified when the axes is forced to lie on a single particle. As expected, with WTA axes, the endpoint of the LO distribution is observed to be $\Nsubnobeta{2,1}=0.5$. With $k_T^{1/2}$ and one-pass $k_T^{1/2}$ axes, the endpoint is extended, as the axes are no longer required to lie on a particle. However, there is still a feature at $\Nsubnobeta{2,1}=0.5$. The fact that the one-pass minimization has a very small effect demonstrates that the seed axes have been appropriately chosen.

In \Fig{fig:endpoint_b_beta2} we consider the behavior of these axes in parton level \vincia{} Monte Carlo, where a large number of partons are present in the final state. As for the case of $\beta=1$, the physical endpoint of the distribution is greatly extended in parton shower Monte Carlo, and the difference between the endpoints for the different axes definitions is also greatly reduced. Note that there is a slight different for $\beta=2$ when WTA axes are used. This is because this defines a legitimately distinct observable.

\begin{figure*}[t]
\centering
\subfloat[]{\label{fig:compare_min_a_beta2_NLO}
\includegraphics[width=8.5cm]{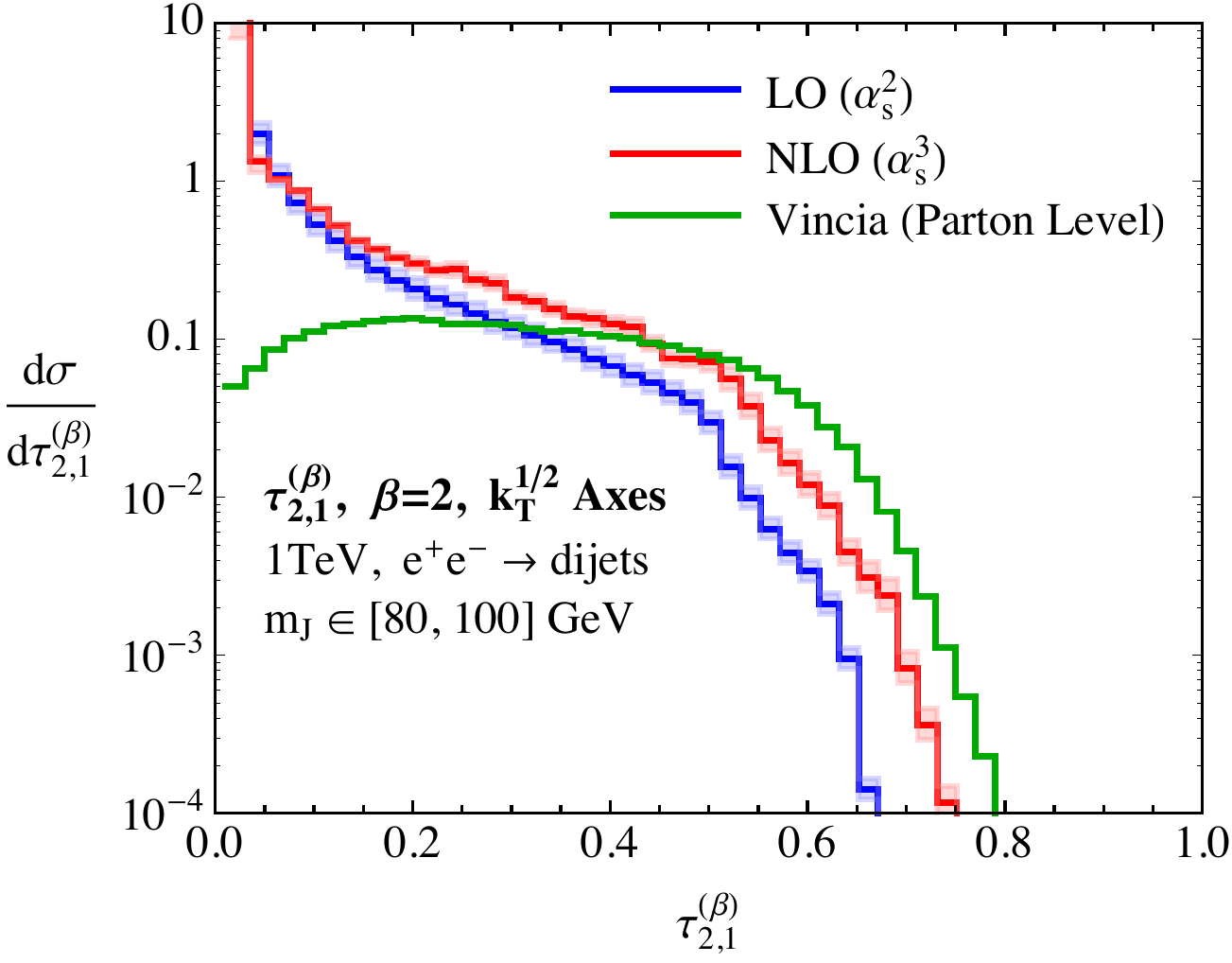}    
} \ \ 
\subfloat[]{\label{fig:compare_min_b_beta2_MC}
\includegraphics[width=8.4cm]{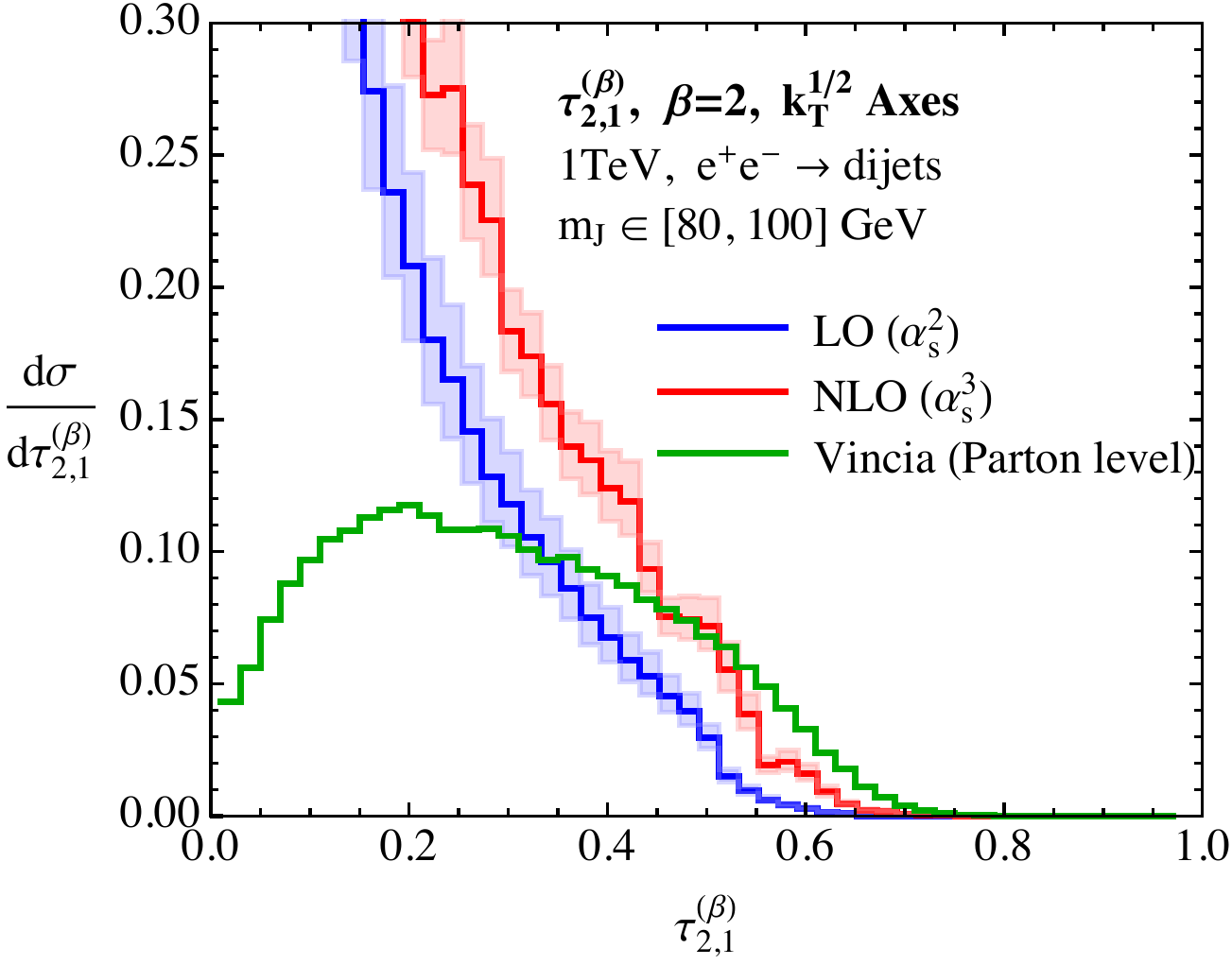} 
}
\caption{Distributions for $\Nsubnobeta{2,1}$ as defined using $k_T^{1/2}$ axes with $\beta=2$ in both fixed order perturbation theory and parton level Monte Carlo. The distributions are shown on a logarithmic scale in (a), and on a linear scale in (b). 
}
\label{fig:wtavop_beta2_2}
\end{figure*}

To understand how the physical regions of the $\beta=2$ distributions are extended at higher orders in perturbation theory, in \Fig{fig:wtavop_beta2_2} we compare the LO, NLO and parton level \vincia{} predictions for $\Nsubnobeta{2,1}$ as defined with  $k_T^{1/2}$ axes and $\beta=2$.\footnote{As for $\beta=1$, similar behavior is observed for  the one-pass minimized axes.}  We have chosen to consider the behavior of the $k_T^{1/2}$ axes, as the WTA $k_T^{1/2}$ axes behave identically to the case of the WTA $k_T$ axes for $\beta=1$ due to the presence of the sharp discontinuity. It is therefore interesting to consider how the NLO behavior is modified when the LO behavior is less drastic, although from \Fig{fig:compare_min_b_beta2_MC}, one can see that there is still a small feature in $\Nsubnobeta{2,1}$ at LO (It is perhaps more visible in the logarithmic plot of \Fig{fig:nsub_sb1}). As with the case of $\beta=1$, we see that the physical region is extended at NLO. There is also a correction at $\Nsubnobeta{2,1}=0.5$, however, much less so than for $\beta=1$, where the LO discontinuity is much larger. In this case, the NLO contribution is also closer to filling out the region explored by the Monte Carlo than for $\beta=1$, implying that perhaps computations could be feasible in this case.

We conclude this section by discussing prospects for the analytic treatment of the $\Nsubnobeta{2,1}$ observable with minimized axes, in particular for the case of $\beta=1$. Having identified the singular feature at $\Nsubnobeta{2,1}=0.5$ one could write down a factorization theorem to resum to all orders the singular logarithms in this region.  However, a factorization theorem which resums singular contributions at $\Nsubnobeta{2,1}=0.5$ merely acts to resum the singular peak right at $\Nsubnobeta{2,1}=0.5$ and must be matched to fixed order outside of the singular region. This implies that it would also have an upper bound of $\Nsubnobeta{2,1}=2/3$.  Since the Monte Carlo distributions go beyond $\Nsubnobeta{2,1}=0.8$, this suggests that extremely high orders in perturbation theory are required to obtain agreement.   However, further confusing the issue is the fact that parton shower Monte Carlo predictions agree well with $\Nsubnobeta{2,1}$ measurements from ATLAS and CMS (see, e.g., \cite{ATLAS:2012am,CMS:2014joa,atlas_recent:2015,Aad:2015rpa}). 

%%%%
%%%%

%%%%
\section{The unresolved limit of $D_2$}\label{sec:unresolved_D2}
%%%%

It is interesting to contrast the behavior of the $\Nsubnobeta{2,1}$ observable with that of $D_2$. A particularly convenient feature of the $D_2$ observable is that it is defined without reference to axes, and therefore has a less subtle behavior in the unresolved limit.

In the presence of a mass cut, $D_2$ regulates all singularities for $D_2>0$. It is instructive to understand how $D_2$ regulates the $\mathcal{O}(\alpha_s^2)$ configurations in \Fig{fig:singular_configs} which gave rise to the divergence at the $\Nsubnobeta{2,1}$ endpoint. For the energy correlation functions evaluated on this configuration, we have parametrically
%%%%
\begin{align}
\ecf{2}{\beta}&\sim z_s +\theta_{cc}^\beta\,, \\
\ecf{3}{\beta}&\sim  z_s \theta_{cc}^\beta\,. \nonumber
\end{align}
%%%%
In particular, unlike for $\Nsubnobeta{2}$, the three point energy correlation function is sensitive to both the splitting angle, $\theta_{cc}$, as well as the energy fraction of the soft parton, $z_s$. This implies that  soft or collinear singularities are inconsistent with the relation $\ecfnobeta{3} \sim \ecfnobeta{2}^2$.

\begin{figure}[t]
\includegraphics[width=8.5cm]{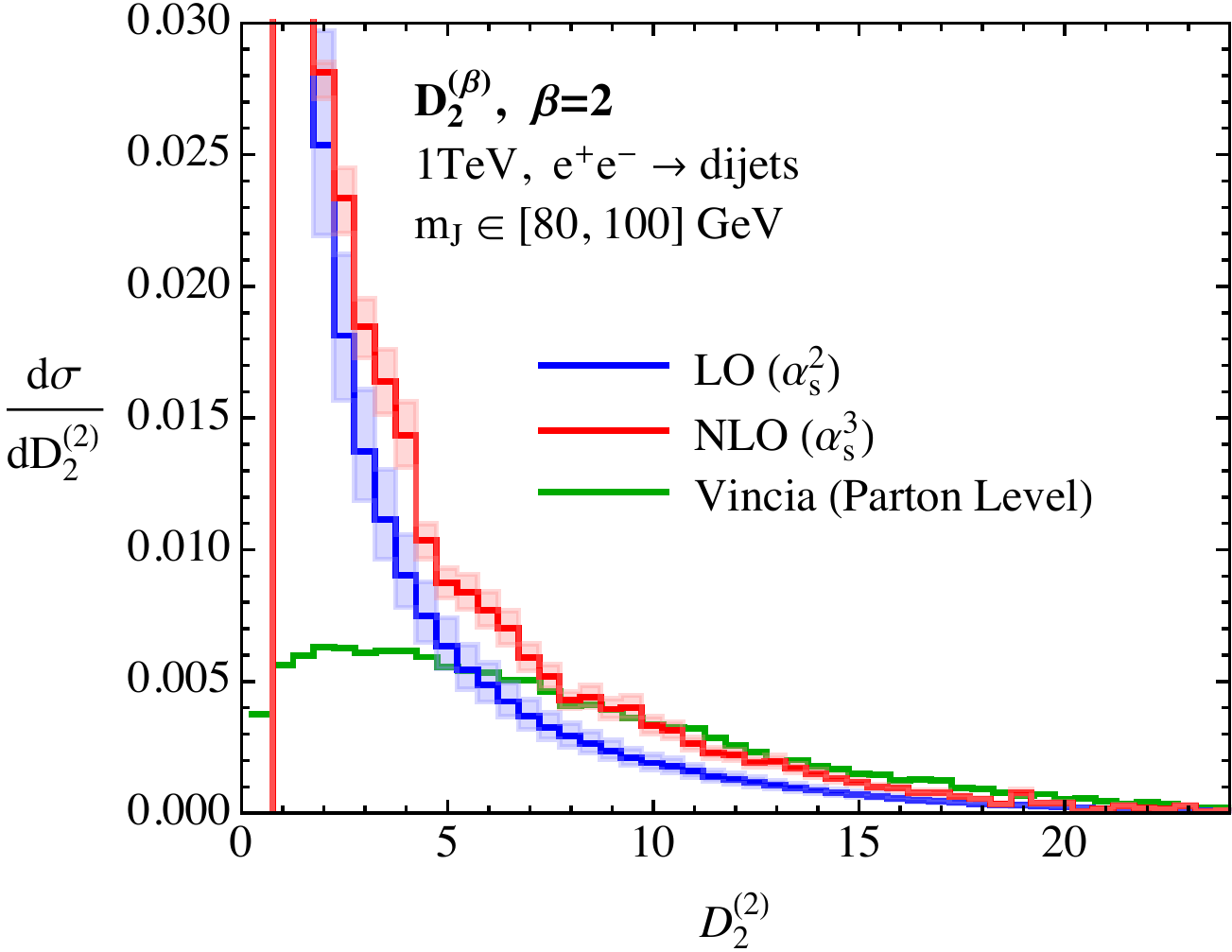} 
\caption{The $D_2$ observable with angular exponent $\beta=2$, as computed in LO and NLO perturbation theory compared with the prediction of the \vincia{} Monte Carlo at parton level.
}
\label{fig:D2}
\end{figure}

In \Fig{fig:D2}, we show the LO and NLO fixed order predictions for the $D_2$ observable, as well as a comparison with \vincia{} parton level Monte Carlo. The singular behavior as $D_2 \to 0$ in fixed order perturbation theory is clearly visible, but no other singularities are present. Good agreement is seen between the shape of the NLO fixed order distribution to the right of the peak, and the resummed result computed from the \vincia{} Monte Carlo. This figure also clearly shows the lack of structure present in the $D_2 \gg 1$ region of phase space.  The fixed order singular structure of $D_2$ has also been discussed in \Ref{Larkoski:2015kga}.

%%%%
%%%%

%%%%
\begin{figure*}[t]
\centering
\subfloat[]{\label{fig:kt_hadr}
\includegraphics[width=7.5cm]{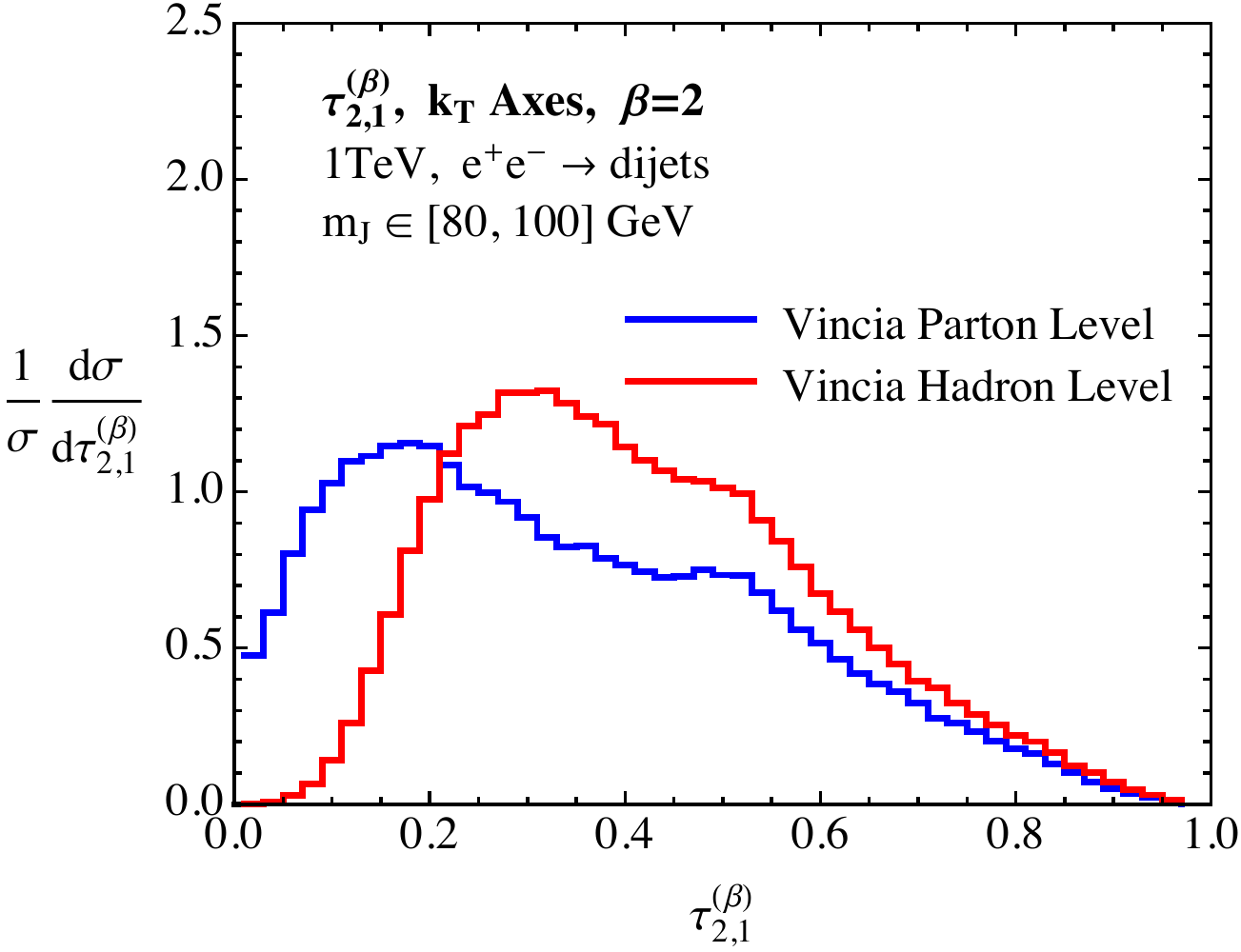}    
} \ \ \hspace{1cm}
\subfloat[]{\label{fig:CA_hadr}
\includegraphics[width=7.5cm]{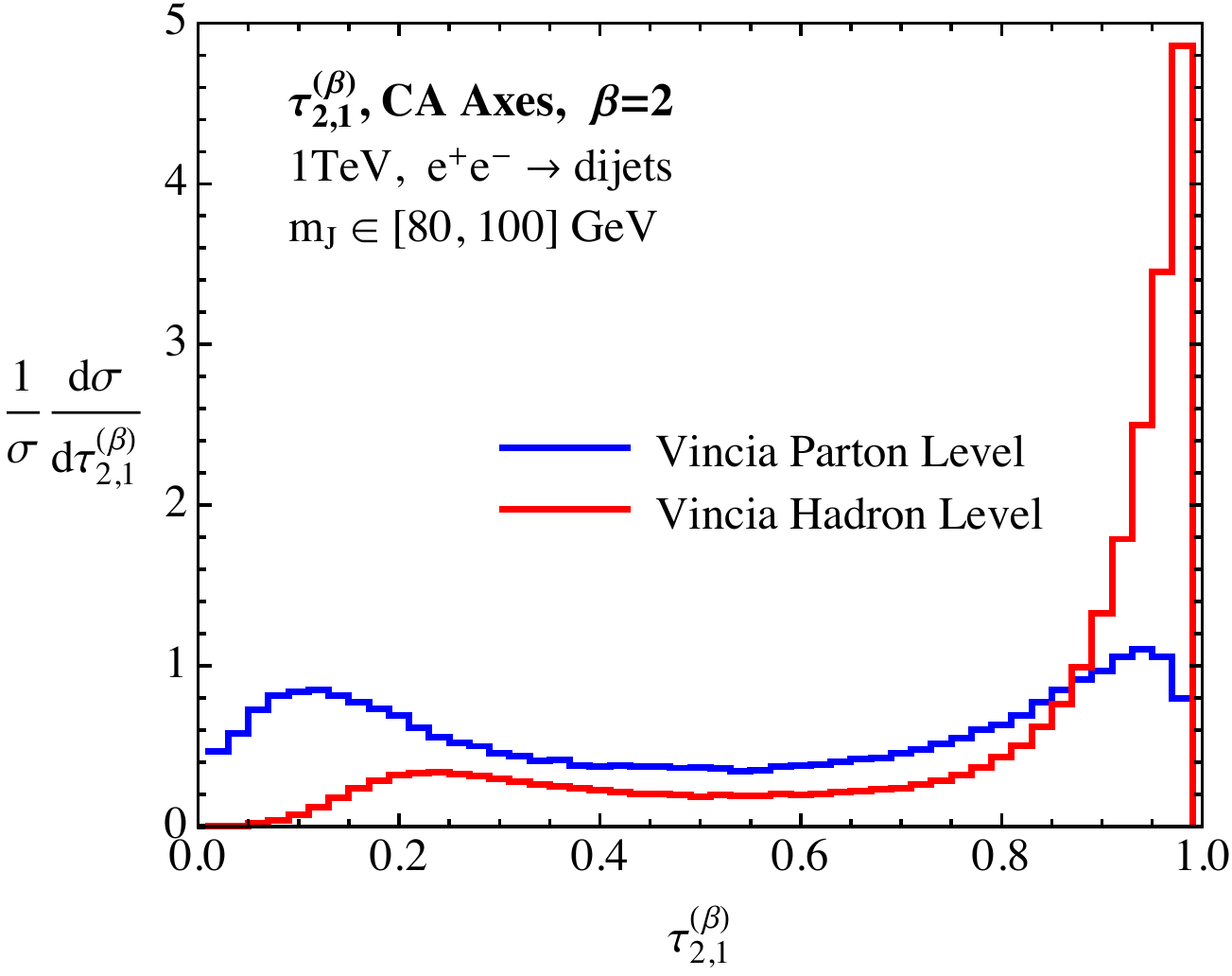} 
}
\caption{A comparison of parton and hadron level Monte Carlo for $\Nsubnobeta{2,1}$ with $\beta=2$. Exclusive $k_T$ axes are used in (a) and exclusive C/A axes are used in (b). Non-perturbative corrections are large at the $\Nsubnobeta{2,1}\to1$ endpoint due to the presence of a soft singularity.
}
\label{fig:tau21_hadr}
\end{figure*}
%%%%

An interesting feature of the $D_2$ observable is that the peak of the perturbative distribution exists in the resolved region, where the two-prong factorization theorems are valid. On the other hand, for many definitions of the $N$-subjettiness axes, the peak of the $\Nsubnobeta{2,1}$ observable is in the unresolved region. Higher order calculations in the unresolved region of phase space are therefore required to even begin to get the correct qualitative shape of the $\Nsubnobeta{2,1}$ distribution.

%%%%
\section{Impact of Hadronization}\label{sec:non_pert}
%%%%

To this point we have discussed the behavior of two-prong substructure observables in a purely perturbative context. However, non-perturbative effects also play an important role in singular regions of phase space,  where the distribution is dominated by emissions characterized by the scale $\Lambda_\text{QCD}$. In this section we emphasize that the variety of singular structures arising for different definitions of the $N$-subjettiness axes implies also a large variation in the impact of hadronization corrections.

For additive jet shape observables the dominant non-perturbative effects can be modeled using  shape functions, which have support over a region of size $\Lambda_{\text{QCD}}$ and are convolved with the perturbative soft function entering the factorized description of a particular singular region \cite{Korchemsky:1999kt,Korchemsky:2000kp,Bosch:2004th,Hoang:2007vb,Ligeti:2008ac}. This approach to incorporating non-perturbative effects analytically is optimal in the case that the non-perturbative contributions do not give large corrections to the shape of the distribution, but instead are well approximated by a shift in the first moment, which in certain cases can be shown to be universal \cite{Akhoury:1995sp,Dokshitzer:1995zt,Dokshitzer:1995zt,Salam:2001bd,Lee:2006fn,Lee:2007jr,Mateu:2012nk}.

When considering non-perturbative corrections to two-prong substructure observables,  one in general needs to incorporate non-perturbative effects through multi-differential shape functions defined in terms of multi-differential factorization theorems \cite{Larkoski:2013paa,Larkoski:2014tva,Procura:2014cba}. This is necessary to incorporate non-perturbative correlations between the observables, for example $\Nsubnobeta{1}$ and $\Nsubnobeta{2}$.  A further complication arises for observables which exhibit multiple singular regions, as a shape function is in principle required for each such region. If the distinct singular regions cannot be well separated, then limited predictive power is retained.

A significant simplification occurs in the resolved limit, where, to leading power, one observable is set by the hard splitting defining the jet substructure. In this case, the dominant non-perturbative effects can be implemented through a single shape function for the remaining variable, which can be given a field theoretic definition within SCET$_+$. Multi-differential shape functions involving correlations are not required. Isolating non-perturbative effects to the resolved region is therefore advantageous for analytic calculations.

To further improve our understanding of the singular structure of both the $\Nsubnobeta{2,1}$ and $D_2$ observables, it is instructive to study the impact of hadronization on their distributions, which we will do using the \vincia{} Monte Carlo. We begin by comparing the impact of hadronization on the $\Nsubnobeta{2,1}$ observable as defined with exclusive $k_T$ and C/A axes, which we used as representative of the different possible perturbative behaviors as $\Nsubnobeta{2,1}\to 1$. Distributions for both axes choices are shown in \Fig{fig:tau21_hadr}  for $\beta=2$. For both axes definitions, large hadronization corrections are observed as  $\Nsubnobeta{2,1}\to 0$. However, while large hadronization corrections are observed for the C/A axes as $\Nsubnobeta{2,1}\to1$, for $k_T$ axes, hadronization has no effect for $\Nsubnobeta{2,1}\gtrsim 0.5$. This is made clear by the feature at $\Nsubnobeta{2,1}\sim 0.5$, which is not smeared or shifted at all by hadronization, demonstrating a rapid transition from non-perturbative to perturbative physics. This behavior clearly reflects the underlying singular structure of the observables. While for C/A axes, non-perturbative effects give an $\mathcal{O}(1)$ contribution throughout the entire distribution, for $k_T$ axes, they are isolated at $\Nsubnobeta{2,1}\to0$, and hence could potentially be well described by a shape function in an SCET$_+$ factorization theorem.

%%%%
\begin{figure*}[t]
\centering
\subfloat[]{\label{fig:min_hadra}
\includegraphics[width=7.5cm]{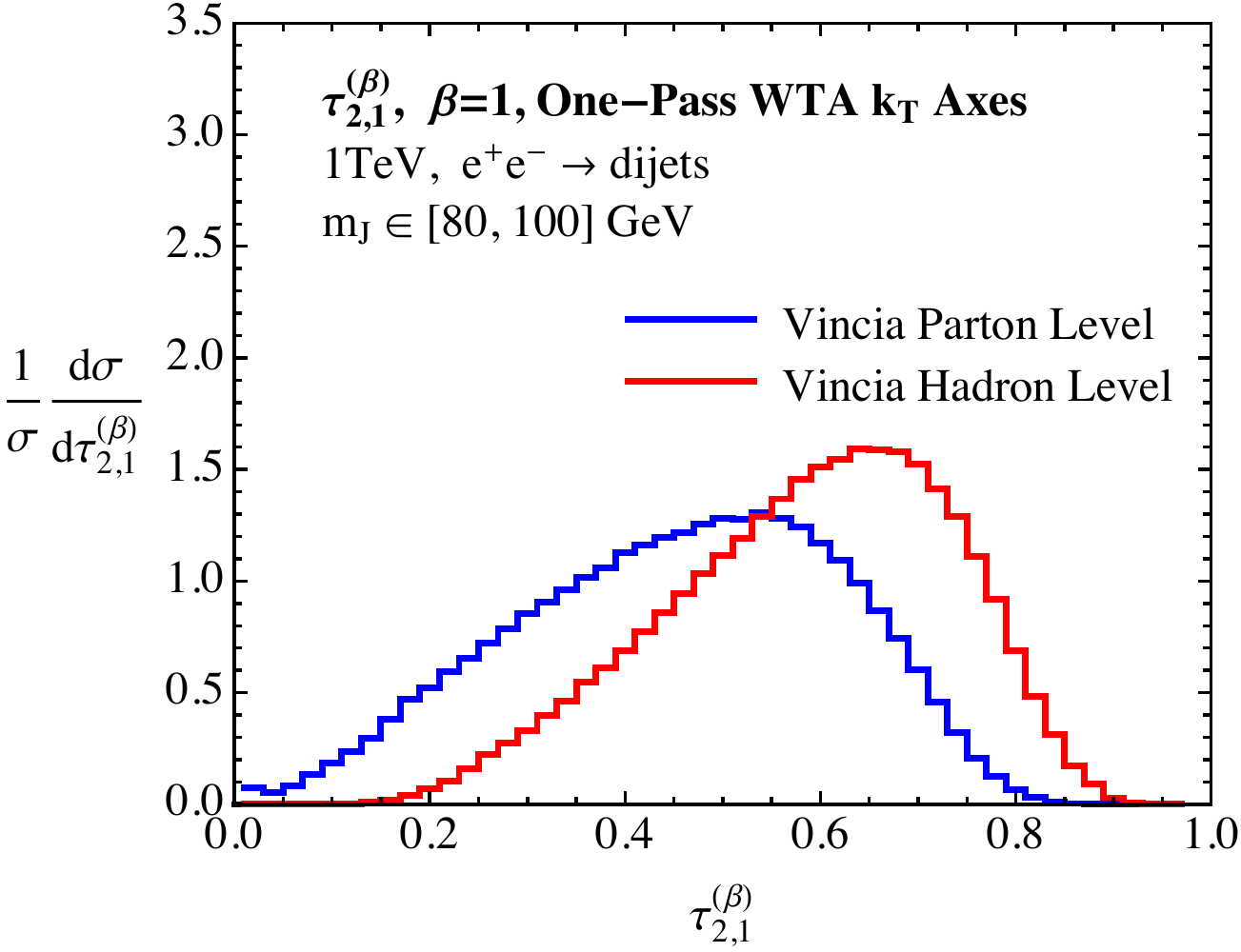}    
} \ \ \hspace{1cm}
\subfloat[]{\label{fig:min_hadra2}
\includegraphics[width=7.5cm]{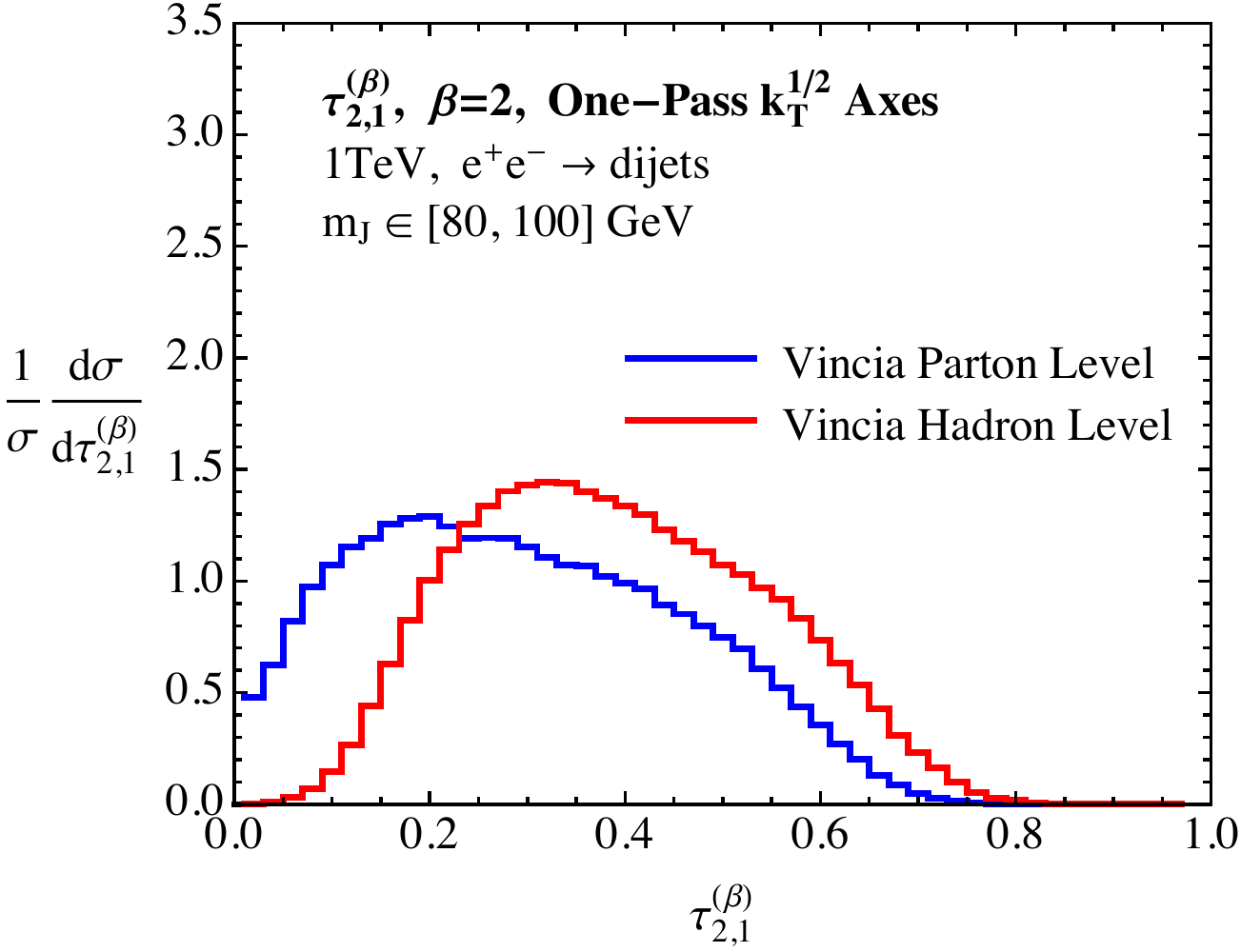} 
}
\caption{Hadronization effects for $\Nsubnobeta{2,1}$ as defined with one-pass minimized WTA $k_T$ axes for $\beta=1$ in (a) and for one-pass minimized WTA $k_T^{1/2}$ axes for $\beta=2$ in (b). Large corrections are observed throughout the entire distribution.
}
\label{fig:min_hadr}
\end{figure*}
%%%%

%%%%
\begin{figure}[t]
\includegraphics[width=7.5cm]{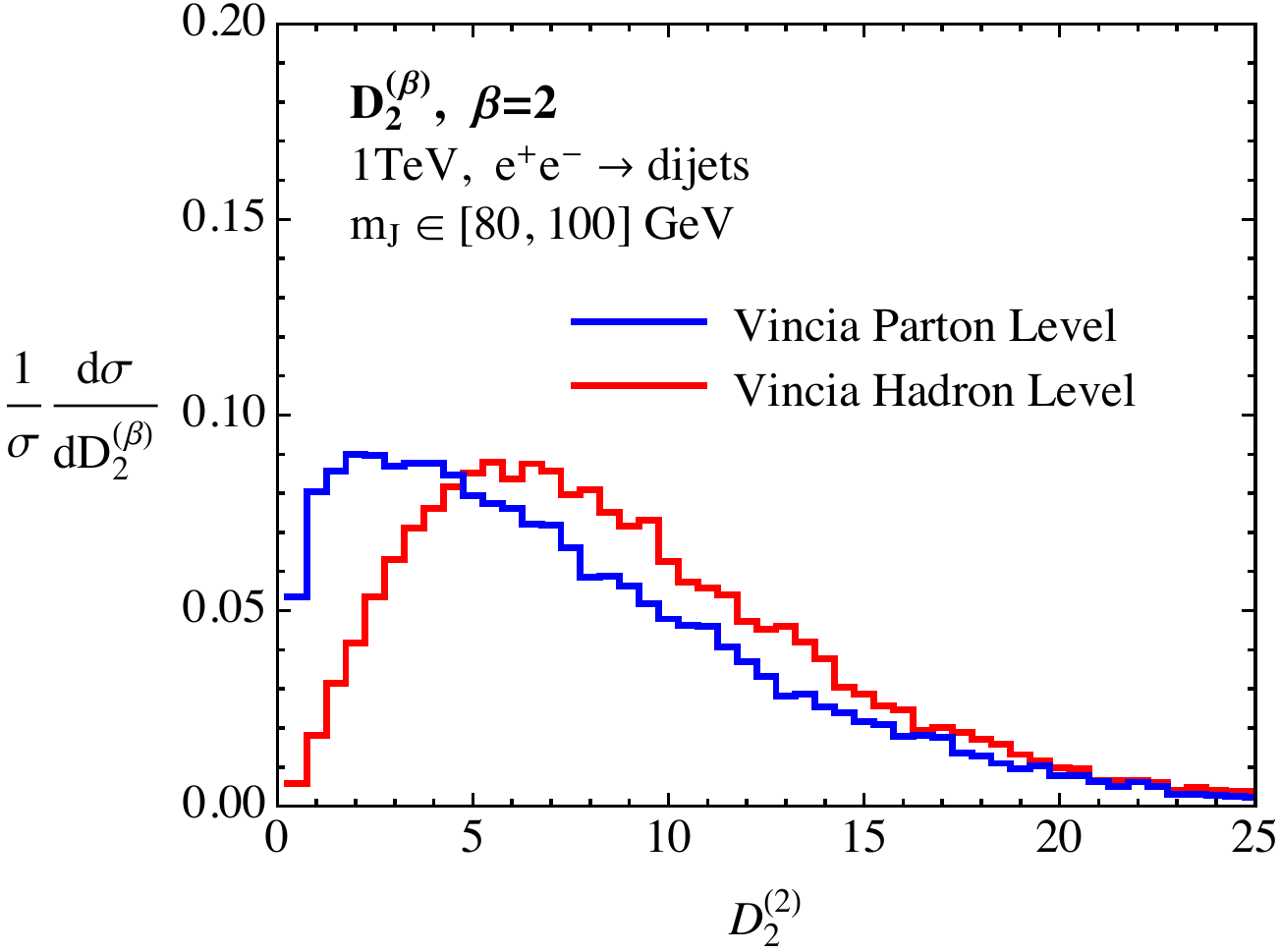}    
\caption{Hadronization effects for the $D_2$ observable with angular exponent $\beta=2$. 
}
\label{fig:D2_hadr}
\end{figure}
%%%%

In \Fig{fig:min_hadr} we compare parton level and hadronized Monte Carlo for $\Nsubnobeta{2,1}$ as defined with one-pass minimized WTA $k_T$-axes with $\beta=1$, and one-pass minimized $k_T^{1/2}$ axes with $\beta=2$, which are the XCone recommended axes \cite{Stewart:2015waa}.\footnote{ As shown in \Figs{fig:endpoint_b}{fig:endpoint_b_beta2} there is little difference in parton level Monte Carlo depending on whether or not the one-pass minimization is performed. This remains true after hadronization.} The effect of hadronization on the distribution is quite interesting. In particular, it is large throughout the entire distribution, representing the fact that there are singularities present throughout the physical region. It is important to emphasize that the non-perturbative effects at different regions in the distribution arise from distinct singularities. A first principles treatment of the non-perturbative effects would require associating distinct non-perturbative shape functions with the factorization theorems for each singularity.\footnote{For the $\beta=1$ case the situation is intrinsically more complicated as there are leading power non-perturbative corrections to both the jet and soft functions. See \Ref{Becher:2013iya} for a discussion. } This presents a serious difficulty to obtaining an analytic calculation of this observable, coupled with the earlier observation that it is unstable in perturbation theory.

Finally, it is worth contrasting the behavior of non-perturbative effects for $\Nsubnobeta{2,1}$ with those for $D_2$, which are shown in \Fig{fig:D2_hadr}. As discussed in \Sec{sec:unresolved_D2}, $D_2$ only has a singularity in the limit $D_2 \to 0$. In this region of phase space, non-perturbative effects are expected to give an $\mathcal{O}(1)$ contribution, as is clearly seen in \Fig{fig:D2_hadr}. For $D_2\gg 1$ non-perturbative effects are minimal. Since non-perturbative effects are isolated in the $D_2 \to 0$ limit, they can be incorporated through an additive shape function. This was studied in \Ref{Larkoski:2015kga} where excellent agreement between a single parameter shape function and Monte Carlo was observed.

%%%%
\section{Conclusions}\label{sec:conc}
%%%%

In this paper we have studied the singular structure of two-prong jet substructure observables, focusing in particular on the behavior of the $\Nsubnobeta{2,1}$ and $D_2$ observables in the unresolved limit. We provided for the first time an understanding of the singular structure of $\Nsubnobeta{2,1}$ throughout the entire available phase space for a massive QCD jet. We have demonstrated that while the structure in the $\Nsubnobeta{2,1} \to 0$ limit can be well understood using universal all orders factorization theorems formulated in SCET$_+$, the structure in the unresolved limit is significantly more complicated, and depends on the definition of the axes used to define the observable. As a consequence of this, non-perturbative corrections to the $\Nsubnobeta{2,1}$ observable also depend on this choice. 

\begin{table}
\begin{center}
\begin{tabular}{c c c | c c c}
 $\beta$ & $d_{ij}$ & Reco. & Tail? & Edge? & Divergence?\\
 \hline
1 & CA & $E$-scheme  & - & - & \checkmark \\
 & CA & WTA  & - & - & \checkmark \\
 & $k_T^{1/2}$ & WTA  & \checkmark & - & - \\
 & $k_T$ & $E$-scheme  & \checkmark & - & - \\
 & $k_T$ & WTA  & - & \checkmark & - \\
 \hline
2 & CA & $E$-scheme  & - & - & \checkmark \\
 & $k_T^{1/2}$ & $E$-scheme  & \checkmark & - & - \\
 & $k_T^{1/2}$ & WTA  & - & \checkmark & - \\
 & $k_T$ & $E$-scheme  & \checkmark & - & - 
\end{tabular}
\end{center}
\caption{
Summary of the behavior at the kinematic endpoint of the fixed-order distributions of $\tau_{2,1}$ for the different observable and axes choices in this paper.  For angular exponents $\beta = 1,2$; the CA, $k_T^{1/2}$ and $k_T$ clustering metrics; and the $E$- and WTA recombination schemes, we show whether the endpoint region has a smooth tail, a sharp edge, or a divergence.
}\label{tab:axes}
\end{table}

Of particular phenomenological interest, we showed that for the case of minimized and WTA axes, which provide the best discrimination power, shoulders appear in the physical region. This implies that high orders in perturbation theory are required to achieve a reasonable description of the observable. The presence of singularities in the physical region also leads to large non-perturbative corrections throughout the entire distribution, making an analytic understanding of the final hadron level distribution difficult. However, for the case of $\beta=2$, it would be interesting to use the understanding of the singular structure elucidated in this paper to perform a perturbative calculation, and to analytically understand the behavior of the resummation at the $\Nsubnobeta{2,1}=0.5$ discontinuity.  In \Tab{tab:axes}, we list the behavior at the kinematic endpoint for the different parameters of $N$-subjettiness studied in this paper.

The behavior of $\Nsubnobeta{2,1}$ was contrasted with that of $D_2$. The $D_2$ observable incorporates three particle correlations, and therefore naturally regulates all soft and collinear singularities away from $D_2=0$. Furthermore, it parametrically separates resolved and unresolved regions of phase space. $D_2$ therefore exhibits several important advantages for precision calculations. It has a simple structure, with a non-singular tail in the unresolved region, and a peak in the resolved region of phase space. Non-perturbative corrections are isolated to the $D_2\to0$ region of phase space, where they can be treated from first principles using a shape function. Indeed, it was these features which allowed a calculation of the $D_2$ observable in \Ref{Larkoski:2015kga}.

Ideally, one would like to extend the level of understanding of two-prong substructure observables to three-prong substructure observables, such as $D_3$ \cite{Larkoski:2014zma} and $\Nsubnobeta{3,2}$ \cite{Thaler:2011gf}, allowing for the analytic study of the substructure of boosted top quark jets. We believe that these observables will manifest many of the same properties discussed in this paper. Indeed it can be expected that the increase to three axes for $\Nsubnobeta{3,2}$ will result in an even more complicated singular structure in the unresolved region. Furthermore, since such observables probe the jet in a more differential nature they are also sensitive to lower scales, and therefore the ability to analytically incorporate non-perturbative corrections will also be important.  If observables relevant for boosted top tagging are to be calculated analytically, it is likely that the choice of an observable with the simplest possible singular structure will play an important role in enabling an analytic treatment of both perturbative and non-peturbative aspects of its calculation.

With jet substructure observables playing a prominent role at the LHC, it is important that they be brought under theoretical control. Analytic calculations are essential to ensure that Monte Carlo programs, which are ultimately used in experimental analyses, accurately reproduce the QCD shower at the level required for substructure observables. Furthermore, if jet substructure variables are to be used to study precision QCD,  a push for precision beyond that which has already been achieved will be required. For such analytic calculations to be practical, it is important that jet substructure observables be designed not only with performance, but also calculational simplicity in mind. One measure of calculability is the complexity of the singular structure of the observables, which was studied in this paper. This emphasis on simplicity is particularly important at early stages, when only a few observables have been analytically calculated. By designing observables with simple singular structures, it is hopeful that this situation can be improved. However, it ultimately seems that for certain substructure observables of interest, a computational approach beyond the usual perturbative expansion in the number of partons is required.  While this is a lofty goal, it would greatly extend the range of observables which could be computed, as well as provide deep insights into the all orders behavior of QCD observables.

%%%%
\acknowledgments
%%%%

We thank Lina Necib, La\'is Schunk, Iain Stewart, Gregory Soyez, and Jesse Thaler for helpful discussions and comments on the manuscript. We thank in particular Duff Neill for many helpful discussions, and collaboration on related topics. We thank Jan Balewski for the use of, and extensive help with, the MIT computing cluster on which part of the fixed-order computations were performed.  This work is supported by the U.S. Department of Energy (DOE) under cooperative research agreements DE-FG02-05ER-41360, and DE-SC0011090.
A.L.~is supported by the U.S. National Science Foundation, under grant PHY--1419008, the LHC Theory Initiative.
  I.M.~is also supported by NSERC of Canada.
Part of the fixed-order computations in this paper were run on the Odyssey cluster supported by the FAS Division of Science, Research Computing Group at Harvard University.

\bibliography{nsub}

\end{document}